\documentclass[final,5p,times,twocolumn,authoryear]{elsarticle}

\usepackage{amssymb}
\usepackage{amsmath}
\usepackage{url}
\usepackage{subfigure}
\usepackage{tabularx}
\usepackage{longtable}
\usepackage{threeparttable}

\journal{Journal of Systems and Software}

\begin{document}

\begin{frontmatter}



\title{Architectural Patterns for Designing Quantum Artificial Intelligence Systems}


\author[inst1]{Mykhailo Klymenko}
\author[inst2]{Thong Hoang}
\author[inst2,inst3]{Xiwei Xu}
\author[inst2,inst4]{Zhenchang Xing}
\author[inst1,inst5]{Muhammad Usman}
\author[inst2,inst3]{Qinghua Lu}
\author[inst2,inst3]{Liming Zhu}

\affiliation[inst1]{organization={Data61, CSIRO},
            addressline={Research Way}, 
            city={Clayton},
            postcode={3168}, 
            state={VIC},
            country={Australia}}

\affiliation[inst2]{organization={Data61, CSIRO},
            addressline={Level 5/13 Garden St}, 
            city={Eveleigh},
            postcode={2015}, 
            state={NSW},
            country={Australia}}

\affiliation[inst3]{organization={University of New South Wales},
            addressline={}, 
            city={Sydney},
            postcode={2052}, 
            state={NSW},
            country={Australia}}

\affiliation[inst4]{organization={Australian National University},
            addressline={}, 
            city={Canberra},
            postcode={2601}, 
            state={ACT},
            country={Australia}}

\affiliation[inst5]{organization={School of Physics, The University of Melbourne},
            addressline={}, 
            city={Melbourne},
            postcode={3010}, 
            state={VIC},
            country={Australia}}

\begin{abstract}
Utilising quantum computing technology to enhance artificial intelligence systems is expected to improve training and inference times, increase robustness against noise and adversarial attacks, and reduce the number of parameters without compromising accuracy. However, moving beyond proof-of-concept or simulations to develop practical applications of these systems while ensuring high software quality faces significant challenges due to the limitations of quantum hardware and the underdeveloped knowledge base in software engineering for such systems. In this work, we have conducted a systematic mapping study to identify the challenges and solutions associated with the software architecture of quantum-enhanced artificial intelligence systems. The results of the systematic mapping study reveal several architectural patterns that describe how quantum components can be integrated into inference engines, as well as middleware patterns that facilitate communication between classical and quantum components. Each pattern realises  a trade-off between various software quality attributes, such as efficiency, scalability, trainability, simplicity, portability, and deployability. The outcomes of this work have been compiled into a catalogue of architectural patterns.
\end{abstract}


%
%
%


\end{frontmatter}


\section{Introduction}
\label{sec:intro}
Exploiting quantum computing technology for artificial intelligence (AI) systems has recently attracted increasing attention as a potential way to enhance their performance and possibly realise quantum advantage~\cite[]{Dunjko_2018, riste_demonstration_2017, SP42M5A9, FKTMJWRS, EBKMYZ93, QV4N5TVV, JA29UXLM}.
The performance enhancement can manifest as faster training, reduced inference time, adversarial or noise robustness, or reduced memory consumption. Quantum advantage refers to the demonstration that a quantum computer can solve a problem that no classical (non-quantum) computer can solve within a feasible amount of time. Establishing or rigorously proving the quantum advantage in quantum AI systems, particularly in quantum machine learning (QML), at their current early stage of development, has become a research topic in itself~\cite[]{harrow_quantum_2017, riste_demonstration_2017, boixo_characterizing_2018, arute_quantum_2019, bravyi_quantum_2020} and remains an open question~\cite[]{terhal_quantum_2018, schuld2022isquantum}. 
Fundamentally, quantum systems provide novel probabilistic models that leverage superposition and entanglement, capabilities that are challenging to achieve on classical computers \cite[]{Preskill2018, broughton2020tensorflow}. These capabilities form the basis for potential quantum advantages.

Numerous studies have explored the physical, mathematical, and algorithmic principles underlying QML; however, only a few works have focused on application designs for existing quantum computers. Specifically, these applications are tailored to Noisy Intermediate-Scale Quantum (NISQ) devices, which are quantum processors with limited qubits and susceptibility to noise, that are available nowadays~\cite[]{Preskill2018}. Despite these constraints, NISQ devices can still offer a valuable platform for developing and testing quantum algorithms in AI applications~\cite[]{bharti2022noisy, coyle2022machine}, addressing various machine learning tasks such as classification and clustering. The inherent limitations of NISQ devices, coupled with non-functional requirements, such as performance, reliability, security, and usability, defined by software quality attributes~\cite[]{BI2021111005, ISO25010}, pose significant challenges in translating quantum algorithms into deployable and high-quality real-world applications.

The challenges outlined above are a central focus of quantum software engineering, which, as outlined in the standard~\cite[]{ISO90003}, aims to utilise engineering principles and methodologies to effectively design, develop, validate, deploy, and evolve software-intensive systems. The creation of high-quality quantum software is a crucial step in the overall progress of quantum computing, as asserted in "The Talavera Manifesto" \cite[]{Piattini2020TheTM}. Quantum software development involves various activities and decision-making processes, which have been reviewed in \cite[]{KHAN2023111682}. Solutions developed during these activities often take the form of formalised architectural and design patterns, providing reusable answers to common challenges in application design \cite[]{bass2003software, Washizaki}. The architectural patterns represent reusable practices and solutions for addressing recurring problems in software architecture~\cite[]{Yue2023}.

Quantum AI systems and quantum software generally share many non-functional quality attributes with traditional software, such as availability, performance, portability, scalability, and compatibility~\cite[]{sodhi2018quality, Yue2023}. However, as established in~\cite[]{Yue2023}, many architectural considerations for quantum AI systems differ from traditional software architectural designs and require a deeper understanding of low-level details. Addressing this challenge necessitates collaborative efforts from software engineers, physicists, and other experts. 

This work aims to identify and analyse existing architectural patterns employed in quantum AI systems.
Specifically, we focus on systems designed for practical applications, solving real-world problems and operating within the constraints of current NISQ hardware. Additionally, we seek to uncover existing trends in the software engineering of quantum AI. Our main contributions to this work are as follows:
\begin{itemize} 
    \item 
    As a result of the systematic mapping study, we identified ten architectural patterns relevant to quantum AI systems: seven related to the quantum-classical split, which is a method for integrating quantum components into these systems, and three related to the quantum middleware layer, which facilitates interactions between quantum and classical systems.
    \item We have investigated the main reasons for integrating quantum computing into AI systems.
    \item Based on the collected publications, we have identified the existing trends in the development of quantum AI systems.
\end{itemize}

\section{Research Methodology}
\label{sec:slr}

\begin{table*}
\caption{List of the primary studies analysed in this work}
\begin{threeparttable}
\footnotesize
\begin{tabular}{| p{.05\textwidth}  p{.7\textwidth} p{.2\textwidth} |} 
\hline
 ID & Title & Ref. \\
\hline
S1 & A classical–quantum convolutional neural network for detecting pneumonia from chest radiographs & \cite{2PPN5V4J} \\
S2 & A Conceptual Architecture for a Quantum-HPC Middleware & \cite{10234288} \\
S3 & A rigorous and robust quantum speed-up in supervised machine learning & \cite{QJA5MGMD} \\
S4 & A shallow hybrid classical–quantum spiking feedforward neural network for noise-robust image classification & \cite{QCVFTLAZ} \\
S5 & Adiabatic quantum linear regression & \cite{TSBCC883} \\
S6 & An artificial neuron implemented on an actual quantum processor & \cite{Tacchino2019} \\
S7 & Application of quantum machine learning using quantum kernel algorithms on multiclass neuron M-type classification & \cite{EBKMYZ93} \\
S8 & Benchmarking adversarially robust quantum machine learning at scale & \cite{PhysRevResearch.5.023186} \\
S9 & Best practices for portfolio optimization by quantum computing, experimented on real quantum devices & \cite{EV88LGAH} \\
S10 & Binding affinity predictions with hybrid quantum-classical convolutional neural networks & \cite{4MJ6LUYI} \\
S11 & Classical-to-quantum convolutional neural network transfer learning & \cite{GSDMNCPI} \\
S12 & Classical–Quantum Transfer Learning for Image Classification & \cite{BVTFZRD6} \\
S13 & Clinical data classification with noisy intermediate scale quantum computers & \cite{6ANRF8NT} \\
S14 & Data Scanning Methods for Quantum-Classical Interface in Quanvolutional Neural Networks & \cite{5NIUVKSK} \\
S15 & Deep quantum neural networks on a superconducting processor & \cite{CP2XZSJH} \\
S16 & Deep Spiking Quantum Neural Network for Noisy Image Classification & \cite{V3QRA9VE} \\
S17 & DeepQMLP: A Scalable Quantum-Classical Hybrid Deep Neural Network Architecture for Classification & \cite{DI9APDS7} \\
S18 & Design of Superior Parameterized Quantum Circuits for Quantum Image Classification & \cite{X25DU5VA} \\
S19 & Development of variational quantum deep neural networks for image recognition & \cite{YGXYUWGG} \\
S20 & Enabling Continuous Deployment Techniques for Quantum Services & \cite{SN5TJDV6} \\
S21 & Enhanced Machine Learning by a Decorated Quantum Circuit & \cite{THZGYRKX} \\
S22 & Enhancing generative models via quantum correlations & \cite{PhysRevX.12.021037} \\
S23 & Entangled topologies for quanvolutional neural networks in quantum image processing & \cite{ULPQR2S6} \\
S24 & Evaluating hybrid quantum-classical deep learning for cybersecurity botnet DGA detection & \cite{KNNZS8DY} \\
S25 & Experimental quantum adversarial learning with programmable superconducting qubits & \cite{9V73AXPJ} \\
S26 & Full-Rotation Quantum Convolutional Neural Network for Abnormal Intrusion Detection System & \cite{X44V646Y} \\
S27 & Hierarchical quantum classifiers & \cite{Grant2018} \\
S28 & High-Dimensional Similarity Search with Quantum-Assisted Variational Autoencoder & \cite{Y4B3NDVJ} \\
S29 & Hybrid Classical-Quantum Artificial Intelligence Models for Electromagnetic Control System Processor Fault Analysis & \cite{DQY64AFH} \\
S30 & Hybrid classical-quantum autoencoder for anomaly detection & \cite{Q5GLRGG8} \\
S31 & Hybrid classical–quantum Convolutional Neural Network for stenosis detection in X-ray coronary angiography & \cite{M2J8JJEZ} \\
S32 & Hybrid Quantum Applications Need Two Orchestrations in Superposition: A Software Architecture Perspective & \cite{9590459} \\
S33 & Hybrid Quantum Machine Learning Assisted Classification of COVID-19 from Computed Tomography Scans & \cite{C75NJPQ6} \\
S34 & Hybrid Quantum Network for classification of finance and MNIST data & \cite{9P945GJK} \\
S35 & Hybrid Quantum-Classical Neural Networks & \cite{86SK6A24} \\
S36 & HyperQUEEN: Hyperspectral Quantum Deep Network For Image Restoration & \cite{4BHTEL8A} \\
S37 & Implementation of Quantum Deep Reinforcement Learning Using Variational Quantum Circuits & \cite{RYNPPVLH} \\
S38 & Implementing a Hybrid Quantum-Classical Neural Network by Utilizing a Variational Quantum Circuit for Detection of Dementia & \cite{T4PG3ASU} \\
S39 & Improved financial forecasting via quantum machine learning & \cite{SP42M5A9} \\
S40 & Integrating Machine Learning Algorithms With Quantum Annealing Solvers for Online Fraud Detection & \cite{NUEYEGIB} \\
S41 & Integration and Evaluation of Quantum Accelerators for Data-Driven User Functions & \cite{4SCZCHNF} \\
S42 & Integration of Classical and Quantum Services Using an Enterprise Service Bus & \cite{759EKCQU} \\
S43 & Introducing Quantum Computing in Mobile Malware Detection & \cite{G7IFR7T9} \\
S44 & Layered Architecture for Quantum Computing & \cite{BKKNMIFR} \\
S45 & Leveraging API Specifications for Scaffolding Quantum Applications & \cite{KXG2SP6W} \\
S46 & Machine learning of high dimensional data on a noisy quantum processor & \cite{ZF8F2Z4B} \\
S47 & Minimizing Deployment Cost of Hybrid Applications & \cite{MWWGF84E} \\
S48 & Network Anomaly Detection Using Quantum Neural Networks on Noisy Quantum Computers & \cite{TYHHK2JT} \\
S49 & On Circuit-Based Hybrid Quantum Neural Networks for Remote Sensing Imagery Classification & \cite{TUNHBV9C} \\
S50 & Parameterized quantum circuits as machine learning models & \cite{Benedetti_2019} \\
S51 & Precise image generation on current noisy quantum computing devices & \cite{NGDQLVU5} \\
S52 & QAmplifyNet: pushing the boundaries of supply chain backorder prediction using interpretable hybrid quantum-classical neural network & \cite{YMMEXC29} \\
S53 & QFaaS: A Serverless Function-as-a-Service framework for Quantum computing & \cite{3V7TMZQ4} \\
S54 & QGFORMER: Quantum-Classical Hybrid Transformer Architecture for Gravitational Wave Detection & \cite{N7LAIMH8} \\
S55 & QSOC: Quantum Service-Oriented Computing & \cite{9RCSA9S3} \\
S56 & Quantum AI simulator using a hybrid CPU–FPGA approach & \cite{NH4H22PM} \\
S57 & Quantum annealing versus classical machine learning applied to a simplified computational biology problem & \cite{2MVBNRUN} \\
S58 & Quantum Annealing-Based Machine Learning for Battery Health Monitoring Robust to Adversarial Attacks & \cite{SGN3AWNT} \\
S59 & Quantum Annealing-Based Software Components: An Experimental Case Study with SAT Solving & \cite{M337EKEK} \\
S60 & Quantum autoencoders for efficient compression of quantum data & \cite{Romero_2017} \\
S61 & Quantum autoencoders with enhanced data encoding & \cite{Bravo-Prieto_2021} \\
S62 & Quantum convolutional neural network for image classification & \cite{BDV87UNJ} \\
S63 & Quantum convolutional neural networks & \cite{NLHULMET} \\
S64 & Quantum convolutional neural networks for multiclass image classification & \cite{CMBTEJF4} \\
S65 & Quantum convolutional neural networks with interaction layers for classification of classical data & \cite{TRCCV6WI} \\
\hline
\end{tabular}
\end{threeparttable}
\label{tab:sources}
\end{table*}

\begin{table*}
\begin{threeparttable}
\footnotesize
\begin{tabular}{| p{.05\textwidth}  p{.7\textwidth} p{.2\textwidth} |} 
\hline
S66 & Quantum discriminator for binary classification & \cite{7LQXEN2W} \\
S67 & Quantum Embedding Search for Quantum Machine Learning & \cite{IS2X25QU} \\
S68 & Quantum Enhancements for AlphaZero & \cite{QV4N5TVV} \\
S69 & Quantum Fourier Convolutional Network & \cite{454HV9MK} \\
S70 & Quantum generative adversarial learning & \cite{PhysRevLett.121.040502} \\
S71 & Quantum Generative Adversarial Networks for learning and loading random distributions & \cite{MBCDDA69} \\
S72 & Quantum generative adversarial networks with multiple superconducting qubits & \cite{LVMXUC3B} \\
S73 & Quantum implementation of an artificial feedforward neural network & \cite{Tacchino_2020} \\
S74 & Quantum Machine Learning for Audio Classification with Applications to Healthcare & \cite{DHEHEQCV} \\
S75 & Quantum Machine Learning in Feature Hilbert Spaces & \cite{PhysRevLett.122.040504} \\
S76 & Quantum Machine Learning with HQC Architectures using non-Classically Simulable Feature Maps & \cite{IXEB9YUY} \\
S77 & Quantum Neural Networks for Resource Allocation in Wireless Communications & \cite{IH8DEBEW} \\
S78 & Quantum neural networks successfully calibrate language models & \cite{5QHTHUXJ} \\
S79 & Quantum semi-supervised generative adversarial network for enhanced data classification & \cite{YN5NEA67} \\
S80 & Quantum Service-Oriented Architectures: From Hybrid Classical Approaches to Future Stand-Alone Solutions & \cite{XUULPYC3} \\
S81 & Quantum service-oriented computing: current landscape and challenges & \cite{Y97WFA6Q} \\
S82 & Quantum Software Architecture Blueprints for the Cloud: Overview and Application to Peer-2-Peer Energy Trading & \cite{D9TYIEZ6} \\
S83 & Quantum Software as a Service Through a Quantum API Gateway & \cite{SRBWSLQR} \\
S84 & Quantum Transfer Learning for Real-World, Small, and High-Dimensional Remotely Sensed Datasets & \cite{10253962} \\
S85 & Quantum Vision Transformers & \cite{Cherrat2024quantumvision} \\
S86 & Quantum-aided secure deep neural network inference on real quantum computers & \cite{JENJXDQR} \\
S87 & Quantum-enhanced deep neural network architecture for image scene classification & \cite{HRPF3VDM} \\
S88 & Quantum-enhanced filter: QFilter & \cite{VKF77E62} \\
S89 & Quanvolutional Neural Networks Powering Image Recognition with Quantum Circuits & \cite{Henderson2020} \\
S90 & QuCardio: Application of Quantum Machine Learning for Detection of Cardiovascular Diseases & \cite{FKTMJWRS} \\
S91 & Reinforcement Quantum Annealing: A Hybrid Quantum Learning Automata & \cite{WBUPYIYH} \\
S92 & Relevance of Near-Term Quantum Computing in the Cloud & \cite{10.1007/978-3-030-72369-9_2} \\
S93 & Review of some existing QML frameworks and novel hybrid classical–quantum neural networks realising binary classification for the noisy datasets & \cite{KMRPKRUT} \\
S94 & Shallow quantum neural networks (SQNNs) with application to crack identification & \cite{LQ7U8YT7} \\
S95 & Shot Optimization in Quantum Machine Learning Architectures to Accelerate Training & \cite{PSVPV246} \\
S96 & Stabilized quantum-enhanced SIEM architecture and speed-up through Hoeffding tree algorithms enable quantum cybersecurity analytics in botnet detection & \cite{7SN73F7Q} \\
S97 & Stereoscopic scalable quantum convolutional neural networks & \cite{5X7N9PDN} \\
S98 & Supervised learning with quantum-enhanced feature spaces & \cite{Q2VWH9J5} \\
S99 & Synergistic pretraining of parametrized quantum circuits via tensor networks & \cite{THVKPQMC} \\
S100 & The Born supremacy: quantum advantage and training of an Ising Born machine & \cite{H43Q8NJS} \\
S101 & The Holy Grail of Quantum Artificial Intelligence: Major Challenges in Accelerating the Machine Learning Pipeline & \cite{RSDKE6EG} \\
S102 & The power of one clean qubit in supervised machine learning & \cite{UG4ZHGN2} \\
S103 & The power of quantum neural networks & \cite{MGV2DIS7} \\
S104 & Toward a Quantum-Science Gateway: A Hybrid Reference Architecture Facilitating Quantum Computing Capabilities for Cloud Utilization & \cite{PQR7YDXZ} \\
S105 & Towards AutoQML: A Cloud-Based Automated Circuit Architecture Search Framework & \cite{4HXFSHKM} \\
S106 & Towards provably efficient quantum algorithms for large-scale machine-learning models & \cite{MZM6BYYV} \\
S107 & Towards quantum enhanced adversarial robustness in machine learning & \cite{JA29UXLM} \\
S108 & Towards quantum machine learning with tensor networks & \cite{Huggins_2019} \\
S109 & Towards Quantum-algorithms-as-a-service & \cite{IVYVVBCE} \\
S110 & Training and Meta-Training Binary Neural Networks with Quantum Computing & \cite{2XPN56TI} \\
S111 & Variational quantum approximate support vector machine with inference transfer & \cite{25YSCYKD} \\
S112 & Variational Quantum Circuits for Deep Reinforcement Learning & \cite{389BSGH3} \\
S113 & When Machine Learning Meets Quantum Computers: A Case Study & \cite{BHB9VE94} \\
\hline
\end{tabular}
\end{threeparttable}
\end{table*}

As our work focuses on architectural patterns related to quantum AI systems, we formulate the following research questions to guide our analysis:

\begin{itemize}
    \item \textbf{RQ1}: \textit{What software architecture solutions are specifically designed for quantum AI systems, what challenges do they address, and what architectural patterns can be identified?}

    \item \textbf{RQ2}: \textit{What are the primary reasons for employing quantum components in AI applications?}
\end{itemize}

\begin{figure} [t!]
    \centering
    \includegraphics[width=\linewidth]{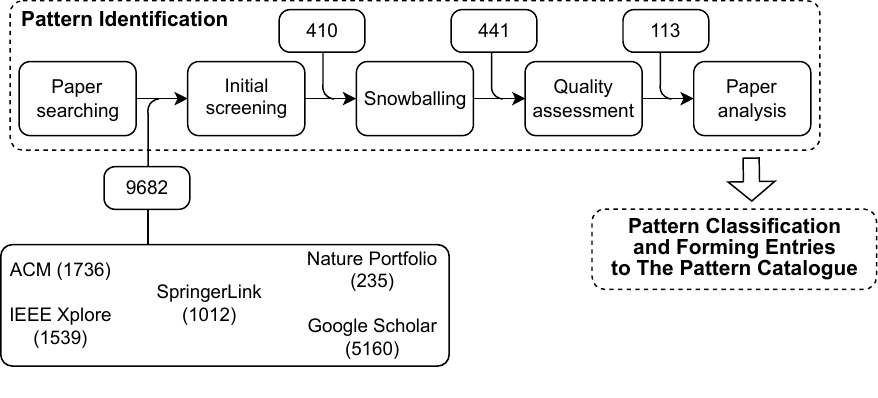}
    \caption{An overall framework for conducting a systematic mapping study}
    \label{fig:slr}
\end{figure}
Figure~\ref{fig:slr} illustrates our evidence-based research methodology for identifying architectural patterns. The methodology is based on a systematic mapping study and is adhering to the guidelines outlined in~\cite[]{Petersen2015}. The framework consists of two main steps: pattern identification and pattern classification. The pattern identification step encompasses several substeps, including paper search, paper collection, snowballing, quality assessment, and paper analysis. We specifically target high-quality academic papers related to architectural patterns in quantum computing systems from electronic databases such as the ACM Digital Library,\footnote{\url{https://dl.acm.org/}} IEEE Xplore,\footnote{\url{https://ieeexplore.ieee.org/Xplore/home.jsp}} SpringerLink,\footnote{\url{https://link.springer.com/}} Nature Portfolio,\footnote{\url{https://www.nature.com/nature-portfolio}} and Google Scholar.\footnote{\url{https://scholar.google.com/}} Broad-scoped scientific databases like GoogleScholar and Scopus can be used interchangeably. While Scopus serves as a general-purpose database in our research methodology, it often retrieves duplicate records, particularly when articles are published in multiple versions or indexed under slightly different author names or journal titles~\cite[]{Liu2021}. This can lead to inefficiencies in the review process, which we aim to minimize. For this reason, we use GoogleScholar to ensure comprehensive coverage while maintaining methodological efficiency. The search string for all databases was composed of two parts, one related to quantum AI and the other to software engineering:

\begin{equation*}
    \parbox{\linewidth}{(``quantum machine learning'' OR ``quantum-classic'' OR ``quantum-classical'' OR ``quantum neural networks'' OR ``quantum convolutional neural networks'' OR ``quantum AI'') AND (``software engineering'' OR ``software architecture'' OR ``architectural patterns'' OR ``architecture'')\\}
\end{equation*}

The collected papers were filtered out based on a series of preset criteria (further details are provided in the protocol available in the supplementary materials). Backward snowballing was employed during the quality assessment phase. In this phase, we analysed references from the primary sources, screened them based on title and context within the citing papers, collected relevant studies, and added them to the list for further quality assessment. Forward snowballing was applied to high-impact papers. We examined the publications citing these sources, screened them by title and abstract, collected relevant studies, and incorporated them into the list for quality assessment. The quality assessment of a publication's reliability and relevance is conducted by answering the following questions with 'yes', 'no', or 'partially':

\begin{enumerate}
    \item Are the research aims clearly stated and focused on quantum AI?
    \item Does the paper demonstrate or provide practical quantum computing applications for AI?
    \item Does the paper explicitly or implicitly include the concept of quantum software architecture?
    \item Does the paper include deployment of existing quantum computing?
    \item Does the paper discuss the advantages of quantum computing?
\end{enumerate}

The first criterion evaluates the scientific integrity and high standards of the research. The second criterion prioritises application-oriented studies in the field of quantum AI. The third criterion assesses the relevance of quantum software engineering. The fourth criterion excludes purely theoretical works that do not consider deployment on quantum computers. The final criterion focuses on papers that demonstrate advantages over traditional systems, highlighting the practical significance of the work. Based on the answers, each publication is assigned a score: 'yes' = 1, 'partially' = 0.5, and 'no' = 0. The scores categorize the papers into three groups: Good (4 $<$ score $\leq$ 5), Fair (2 $<$ score $\leq$ 4), and Poor (0 $<$ score $\leq$ 2). Papers scoring higher than 2 are considered good or fair and included in the review. Those scoring 2 or below are considered irrelevant papers. Figure~\ref{fig:slr}  provides a summary of the number of selected papers at each stage of the systematic mapping study.

After conducting a quality assessment, we selected 113 papers to form the list of primary studies (see Table~\ref{tab:sources}). The collected papers are analyzed in the following steps: gathering metadata such as titles, authors, and full BibTeX records; collecting findings from various studies that address the research questions; and classification the identified architectural patterns based on those findings (the full extraction form and the list of the selected primary sources are provided in the supplementary materials).

The sources listed in Table~\ref{tab:sources}) are relevant to research questions RQ1 and RQ2 and are analysed in Sections 4 and 5, respectively, to address them. 

\section{Reference architecture, role of architectural patterns in quantum AI and related works}
\label{sec:method}

\begin{figure*} [t!]
    \centering
\includegraphics[width=0.8\textwidth]{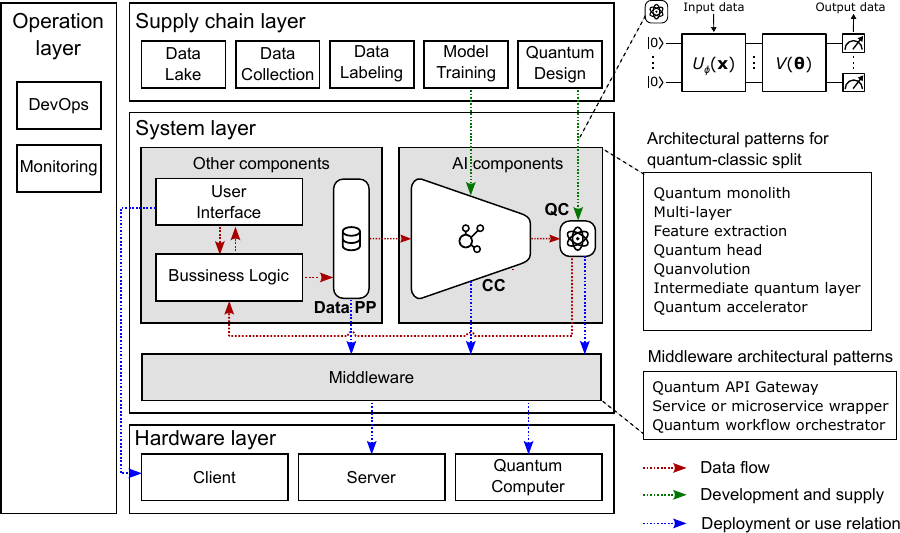}
    \caption{Reference architecture for quantum-enhanced AI systems, featuring data pre-processing (Data PP), classical components (CC), and quantum components (QC), annotated with relevant architectural patterns.}
    \label{fig:arch}
\end{figure*}

\begin{figure} [t!]
    \centering
    \subfigure[]{\includegraphics[width=0.45\linewidth]{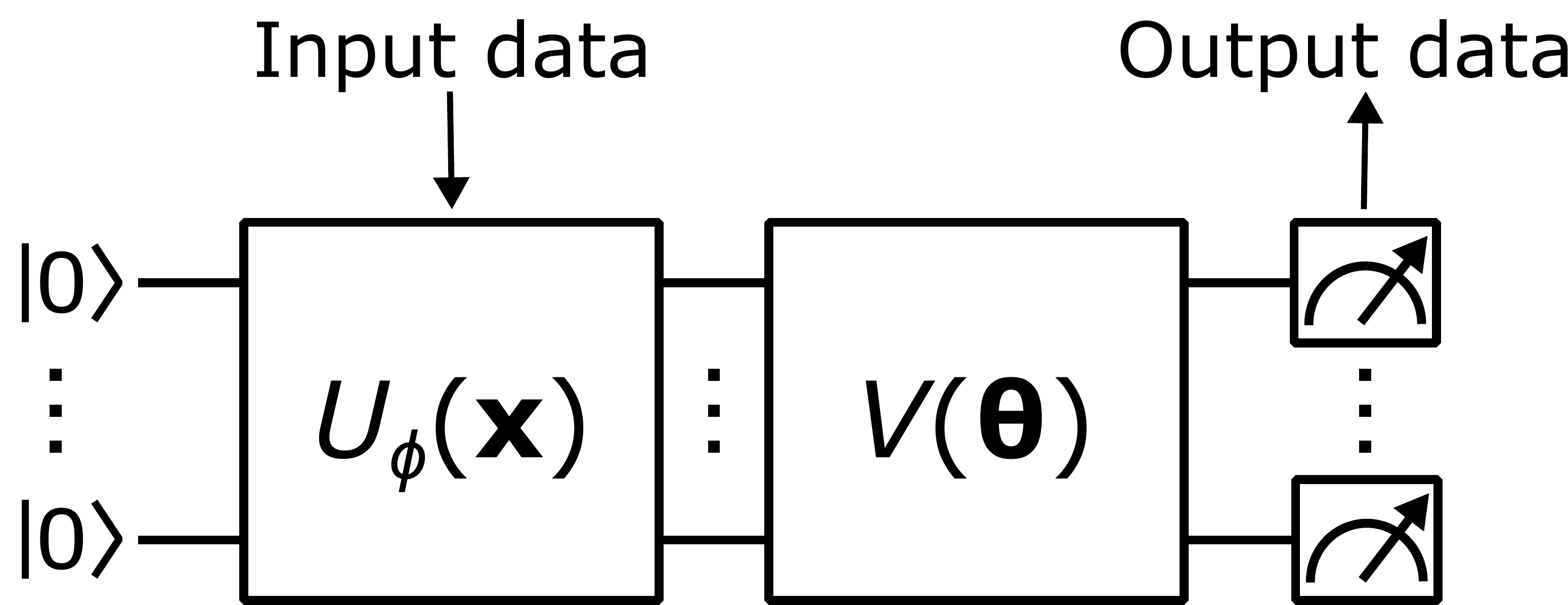}}\\
    \subfigure[]{\includegraphics[width=0.75\linewidth]{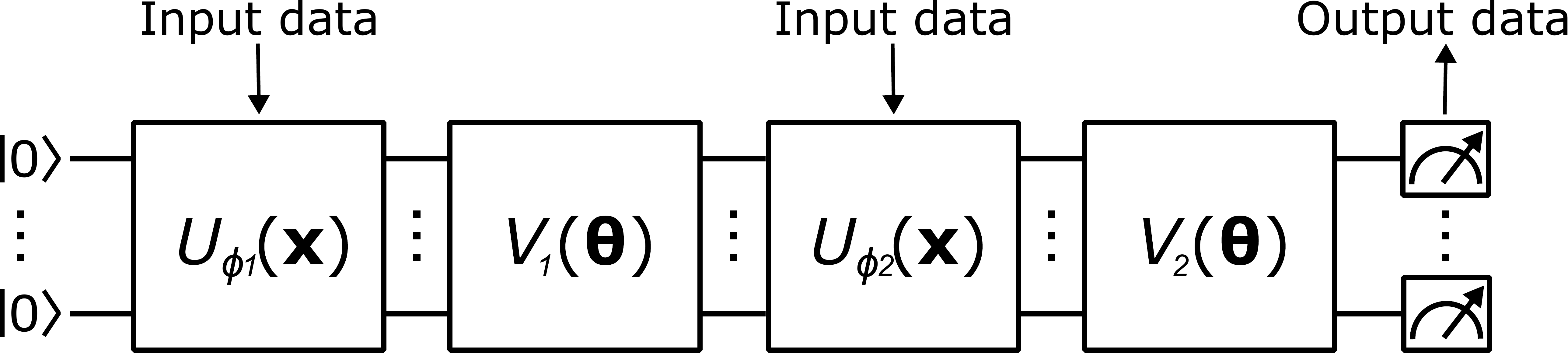}}\\
    \subfigure[]{\includegraphics[width=0.45\linewidth]{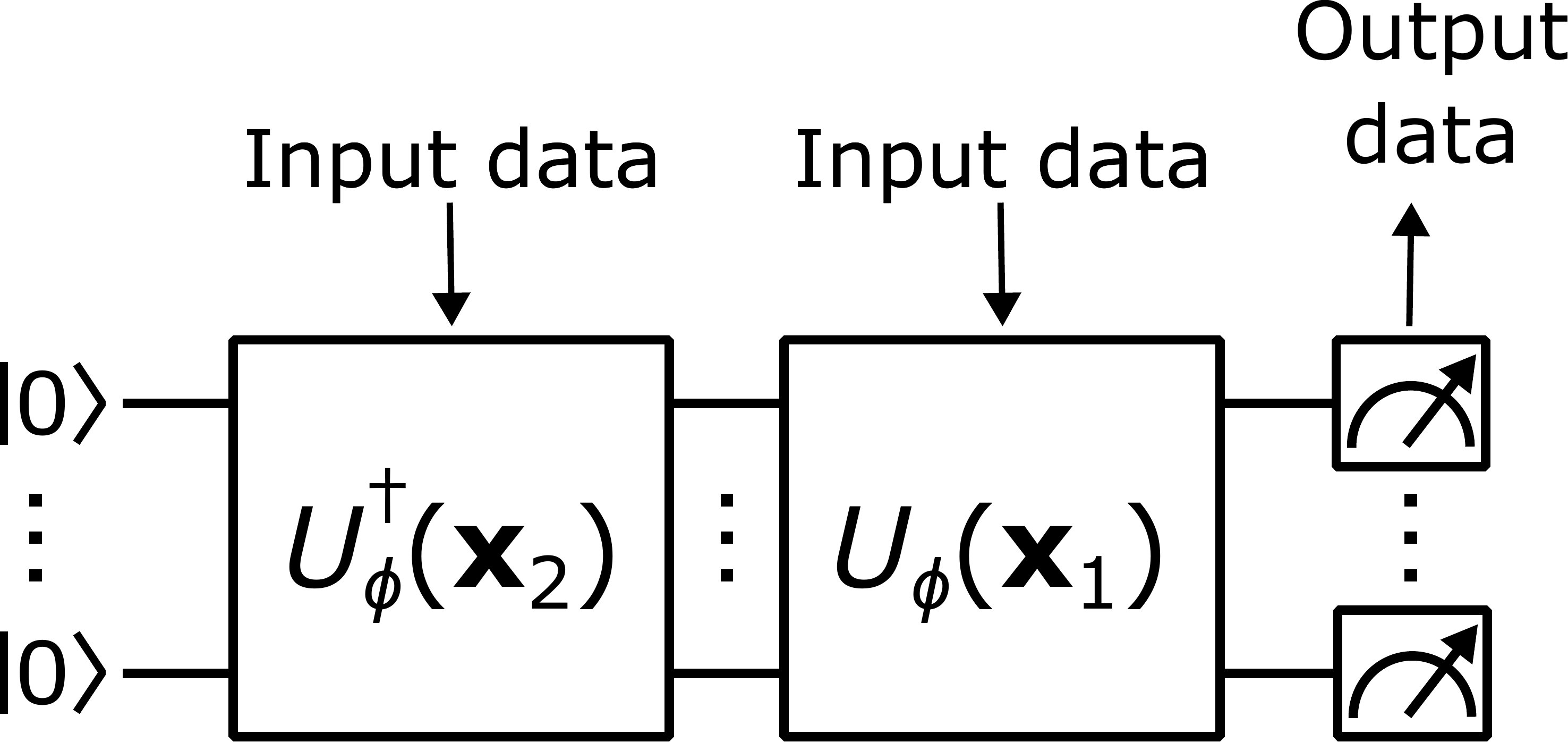}}
    \subfigure[]{\includegraphics[width=0.65\linewidth]{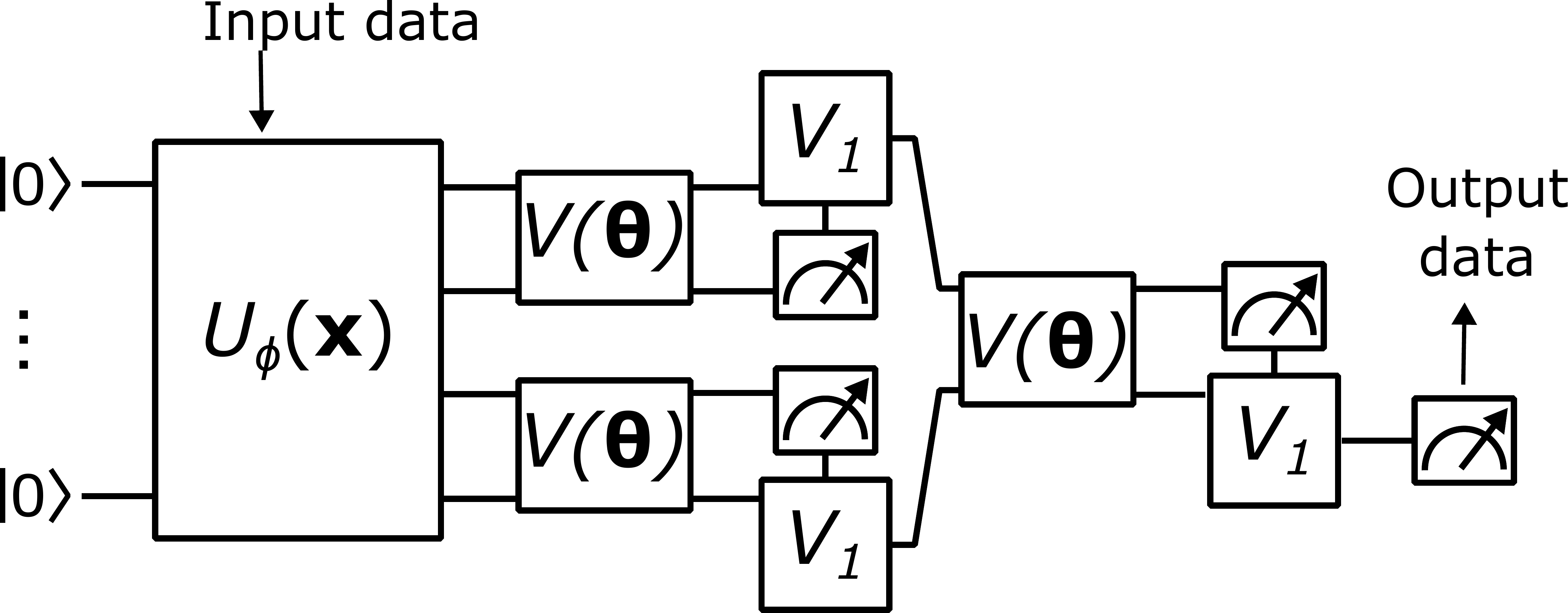}}
    \subfigure[]{\includegraphics[width=0.7\linewidth]{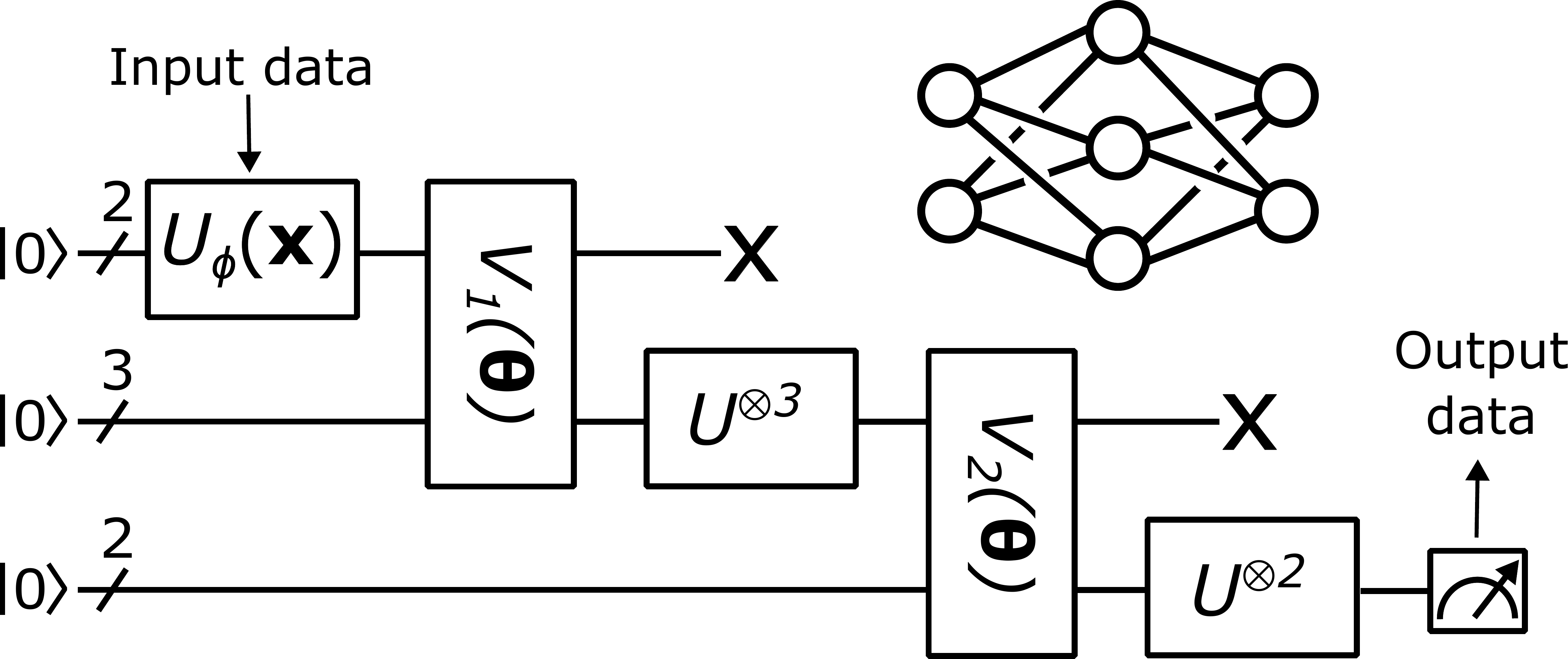}}
        \caption{Quantum circuits used in quantum machine learning applications: a) variational quantum circuit, b) variational quantum circuit implementing data re-uploading \cite[]{KMRPKRUT} c) quantum kernel of the support vector machine \cite[]{QJA5MGMD}, d) quantum convolutional neural network \cite[]{NLHULMET}, and e) deep quantum neural network \cite[]{Beer2020, beer2021training}.}
    \label{fig:cirquit}
\end{figure}

Figure~\ref{fig:arch} illustrates a multi-layered reference architecture for a quantum-enhanced AI system, highlighting various architectural patterns. The architecture is similar to the one described by \cite[]{10553223} 
 and is divided into several key layers, each playing a crucial role in the overall design of the quantum AI system. We present these layers as follows:

\noindent \textbf{Operation layer.} This layer in the architecture is responsible for the ongoing management, maintenance, and monitoring of the AI system, ensuring its smooth and efficient functioning. The two main components of this layer are described as follows: 

\begin{itemize}
    \item \textit{DevOps:} This component integrates development and operations processes to streamline the software development lifecycle and ensure continuous delivery and integration.
    \item \textit{Monitoring:} It ensures the continuous tracking and assessment of the quantum AI system's performance. 
\end{itemize}

\noindent \textbf{Supply chain layer.} This layer encompasses various components essential for managing and optimizing supply chain operations using both classical and quantum computing techniques. These components collaborate to collect, process, and analyze data, ultimately enhancing decision-making and efficiency in supply chain management. We present the components as follows:
\begin{itemize}
    \item \textit{Data lake:} Stores of structured and unstructured data. 
    \item \textit{Data collection:} Gathers data from various sources. 
    \item \textit{Data labeling:} Annotates data for machine learning tasks. 
    \item \textit{Model training:} Employs classical and quantum engines to facilitate the training of AI models. 
    \item \textit{Quantum design:} Involves the development of quantum algorithms and quantum circuits. 
\end{itemize}

Here, we briefly outline the most common examples of quantum circuit architectures that may arise as a result of the quantum software design process for AI systems. Figure~\ref{fig:cirquit}a represents a common and general design for variational quantum circuits. A quantum circuit is one of the models of quantum computing serving as a representation of a quantum program. The circuit describes transformations of quantum states of qubits, which are depicted as horizontal lines in Figure~\ref{fig:cirquit}.  At the beginning of the program, all qubit states are typically initialised to a ket vector $\vert 0 \rangle$, which represents a vector, expressed using Dirac notation, in a complex vector space \cite[]{Nielsen_Chuang_2010}. The input data, represented by the vector $\mathbf{x}$, is encoded into a quantum state via the parameterised unitary operator $U_{\phi}$. There are a variety of possible designs for this unitary operation and corresponding circuitry characterised by different time complexity and robustness. Among them are basis encoding, amplitude encoding, angle encoding, and IQP-style encoding \cite[]{weigold2021encoding, larose2020robust, rath2023encoding}. The prepared quantum state is then modified by a circuit implementing another unitary operation $V(\boldsymbol\theta)$, which contains learnable parameters $\boldsymbol\theta$. The unitary operations $U_{\phi}$ and $V(\boldsymbol\theta)$ consist of rotational single qubit gates and multiqubit controlled gates that introduce quantum entanglement~\cite[]{Nielsen_Chuang_2010}. The learnable parameters $\boldsymbol\theta$ are usually rotation angles in the single-qubit rotational gates. The output is obtained by measuring the quantum states, which converts the quantum states back into classical data. Data encoding does not need to be implemented just once. One of the design solutions employed alongside the variational quantum circuit is a so-called data re-uploading \cite[]{KMRPKRUT}. Data re-uploading is a subclass of quantum embedding implemented repeatedly in several quantum layers, as is illustrated in Fig.~\ref{fig:cirquit}b. This approach can enhance both the expressivity and trainability of the model. The circuit in Fig.~\ref{fig:cirquit}c presents a quantum algorithm that computes the kernel function for the implementation of a quantum-kernel support vector machine. It computes a feature map based on the distance between two data points, $\mathbf{x}_1$ and $\mathbf{x}_2$. The circuit in Fig.~\ref{fig:cirquit}d illustrates the design of a quantum convolutional neural network as proposed in \cite[]{NLHULMET}. This neural network includes analogs of classical pooling layers, which are implemented using the unitary gates $V_1$. These gates have parameters that are determined by measurements from other qubits. The circuit in Fig.~\ref{fig:cirquit}d represents the design of a quantum deep neural network~\cite[]{Beer2020, beer2021training}. In this design, each layer of the quantum neural network, depicted from left to right in the inset in Fig.~\ref{fig:cirquit}e, corresponds to a multi-qubit channel arranged from top to bottom in the circuit diagram.

\noindent \textbf{Hardware layer.} This layer includes computational resources necessary for implementing quantum AI systems. The main components of the hardware layers are described as follows: 
\begin{itemize}
    \item \textit{Client:} This component represents an end-user or a user's device that interacts with quantum AI systems. 
    \item \textit{Server:} The server aims to process user requests. 
    \item \textit{Quantum computer:} It executes quantum algorithms and performs computations in quantum AI systems. 
\end{itemize}

\noindent \textbf{System layer.} This layer of the quantum AI system architecture is responsible for executing the trained models and making predictions or decisions based on input data. The \textit{architectural patterns} identified in this work belong to this layer and offer solutions to the challenges associated with integrating quantum and classical components, ensuring modularity, scalability, and efficient collaboration between various layers. The quantum components can be represented by quantum circuits, with design solutions illustrated in Figure~\ref{fig:cirquit}. In this work, we have identified the following architectural patterns, described in detail in Section \ref{sec:usage}: 

\begin{itemize}
    \item \textit{Quantum middleware architectural patterns:} Middleware architectural patterns facilitate the interaction between quantum and classical components by managing and routing requests and supporting the development of small, independent quantum services.
    \begin{itemize}
        \item Quantum API gateway: Manages and routes requests between quantum and classical components.
        \item Service or microservices wrapper: wraps quantum components into services or microservices.
        \item Quantum workflow orchestration pattern: Detects and orchestrates quantum workflows in the quantum software.
    \end{itemize}
    \item \textit{Quantum-classic split architectural patterns:} Quantum-specific architectural patterns focus on integrating quantum components for data processing and accelerating computations with dedicated quantum hardware. In the presented in Figure~\ref{fig:arch}  architecture, the quantum inference engine is connected sequentially after the classical inference engine. This is only one of the many possible ways to integrate a quantum component. We call this pattern the "Quantum Head". In the course of the systemtic mapping study, we have identified other alternative patterns, such as:

    \begin{itemize}
        \item Quantum monolith: A single, unified quantum component.
        \item Multi-layer: Integrates quantum processing as multiple layers of the architecture.
        \item Quantum feature engineering: Quantum components that handle the initial stages of data processing and implement methods for extracting features from data.
        \item Quantum head: A quantum component that handles the last stages of the inference process.
        \item Quanvolution: A quantum analog of convolutional layers in neural networks.
        \item Intermediate quantum Layer: A quantum layer situated between classical layers.
        \item Quantum accelerator: Hardware specifically designed to accelerate numerical computations.
    \end{itemize}
\end{itemize}

Note that several publications have already addressed the identification of architectural and design patterns in quantum software \cite[]{leymann2019towards, weigold2021hybrid, weigold2021encoding, 9T8ZKDNT, khan2022software}, including the proposal of a specialised pattern language \cite[]{leymann2019towards}. While these works focus on quantum software in general, this paper specifically concentrates on architectural patterns relevant to quantum AI and quantum machine learning.

\section{Pattern catalog for quantum AI systems (RQ1)}
\label{sec:usage}

This section provides a detailed overview of the architectural patterns identified during the systematic mapping study.  Each pattern presents a high-level solution to a challenge encountered in the design, implementation, and management of AI systems \cite[]{2023responsible}. In this work, similar to \cite[]{10.1145/3626234}, each pattern is formally represented using a template that includes the following components: name (with an optional graphical representation), summary, description of the associated problem, solution provided by the pattern, benefits, drawbacks, and known uses. In this work, we classify architectural patterns specific to quantum AI systems into two categories: quantum-classical split patterns (discussed in a subsection \ref{sec:sp}) and quantum middleware patterns (presented in a subsection \ref{sec:mp}).

\subsection{Quantum-classic split architectural patterns (SP)}
\label{sec:sp}

All quantum-classical split patterns are defined within the context of hybrid classical-quantum AI systems. In quantum computing, the hybrid quantum-classic algorithm represents a specific design solution, formalised as the design pattern known as the quantum-classic split~\cite[]{leymann2019towards, weigold2021hybrid}. Within quantum AI systems, real-world applications typically process classical data and produce classical outputs with few exceptions. One of such exceptions is simulating quantum objects, such as molecules and electronic systems~\cite[]{smith2019simulating, JKHCVY9X, cheng2020application}. Additionally, some experimental setups operate exclusively on quantum data, performing classification of quantum states, which is crucial for validating quantum principles and exploring new algorithms~\cite[]{PhysRevX.12.021037, MBCDDA69, Grant2018}. For practical AI software applications, a hybrid quantum-classical software architecture is essential and, arguably, necessary. Our systematic mapping study confirms this, as all 113 selected papers on quantum AI systems discuss hybrid quantum-classical software architectures. In all cases, the extent of work delegated to the classical part versus the quantum part is a critical system-level architectural decision. Therefore, this work treats the quantum-classic split as a family of architectural patterns rather than a single design pattern. It is important to note that the quantum-classic split constitutes a unique family of architectural solutions exclusive to quantum software, with no equivalents in classical software engineering.

\begin{table*}[!ht]
    \centering
        \caption{Quantum-Classic Split Architectural Patterns}
        \footnotesize
\begin{tabularx}{\textwidth} { 
  | >{\raggedright\arraybackslash}p{\dimexpr.15\linewidth-2\tabcolsep-1.3333\arrayrulewidth} 
  | >{\raggedright\arraybackslash}p{\dimexpr.6\linewidth-2\tabcolsep-1.3333\arrayrulewidth}
  | >{\raggedright\arraybackslash}p{\dimexpr.25\linewidth-2\tabcolsep-1.3333\arrayrulewidth} | }
 \hline
 Name of the pattern (ID)  & Summary & Refs. \\ 
 \hline Quantum Monolith (SP-1) & 
This pattern describes the direct encoding of input classical data into quantum states, which are then processed by a quantum computer. The output is generated by the quantum computer as a result of quantum state measurements. The inference task is fully delegated to a quantum component, without any involvement of classical components.  & [S6; S8-9; S15; S22; S25; S27; S37; S50-51; S60-61; S63-66; S70-73; S77; S79; S91; S99-100; S103; S108; S112;] \\ 
\hline Multi-layer (SP-2) & Incorporating multiple quantum components with learnable parameters, interconnected through classical channels, can significantly enhance the overall performance of an AI system by improving expressiveness and accuracy. & [S6; S17; S35; S73; S110; S111] \\
 \hline Quantum Feature Engineering  (SP-3)& This pattern applies when feature engineering is delegated to a quantum computer, which serves as the first step in the inference pipeline by operating directly on the input data. The classical inference engine then carries out subsequent steps in the pipeline. & [S3; S7; S13; S19; S41; S46; S56; S68; S75-76; S87; S93; S98; S102; S111;]\\ 
 \hline 
Quantum Head  (SP-4)& This pattern implements an architecture where a quantum inference engine operates in the pipeline after a classical neural network. The classical component is employed to reduce the dimensionality of the input data.  
& [S1; S11-12; S18; S21; S33; S31; S38; S48; S52; S54; S67-68; S78; S84; S94; S107;  S111] \\
\hline
Quanvolution (SP-5)& This architectural solution replaces the first several layers in classical convolutional neural networks by parameterised quantum circuits operating in parallel or sequentially. The circuit has a relatively small number of qubits and operates as a convolutional filter.  & [S10; S14; S23; S26; S43; S62; S69; S74; S88-90; S97] \\
\hline
 Intermediate Quantum Layer  (SP-6)& Similar to the Quantum Head pattern, this pattern involves a pipeline of classical and quantum inference engines. In this approach, classical components are employed in both the initial and final stages of the inference pipeline, while the quantum component is used in the intermediate stage.  & [S4; S16; S24; S28-30; S33-34; S36; S39; S49; S93; S95; S113]\\
\hline
Quantum Accelerator (SP-7)& This pattern employs a quantum component to evaluate a specific, well-defined function within the system. This quantum component, called a quantum accelerator, usually has a classical analog with lower performance, introduces dependencies on operations executed by a quantum computer, and typically does not possess trainable parameters. & [S5; S40-41; S57-59; S85-86; S101; S106; S113]\\
    \hline
\end{tabularx}
    \label{tab:split}
\end{table*}

We have identified seven architectural patterns related to the quantum-classical split, as compiled in Table~\ref{tab:split}. These patterns address the integration of quantum components within a classical inference engine in the field of quantum AI. The table provides the proposed names of these patterns, their summaries, and a complete list of primary sources, capturing all known instances of their use. A comprehensive description of each identified quantum-classic split pattern is provided below, along with three representative use cases for each pattern.

\noindent \textbf{\textit{SP-1: Quantum monolith.}}

\begin{figure}[h!]
    \centering
    \includegraphics[height=0.18\linewidth]{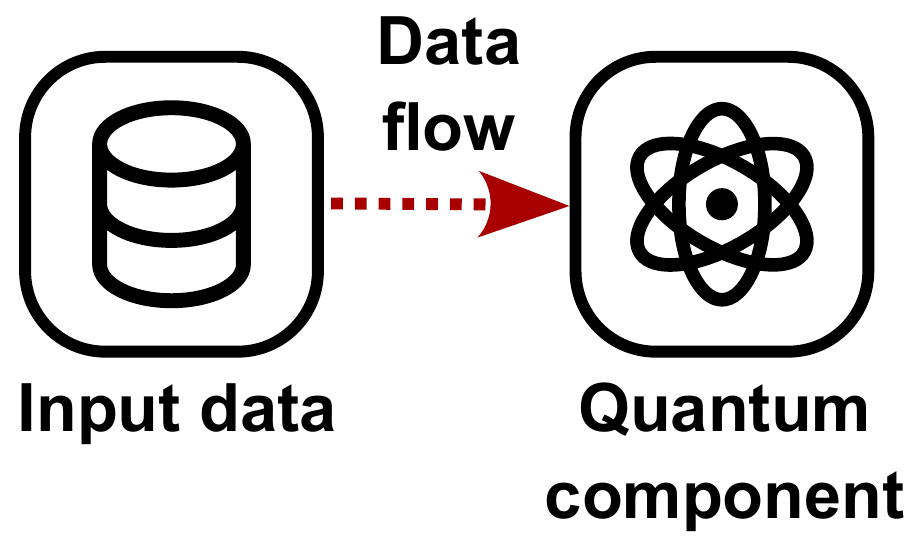}
    \caption{Quantum monolith}
    \label{fig:monolith}
\end{figure}

\noindent \textit{Summary:} This pattern describes the direct encoding of input classical data into quantum states, which are then processed by a quantum computer. The output is generated by the quantum computer as a result of quantum state measurements. The inference task is fully delegated to a quantum component, without any involvement of classical components. Figure~\ref{fig:monolith} provides a simple graphical representation of the quantum-monolith pattern.

\noindent \textit{Problem:} This pattern is designed to enhance AI system performance by maximising the quantum advantage provided by quantum computers.

\noindent \textit{Solution:} In this approach, the quantum computer is solely responsible for inference, while the classical computer handles only the pre-processing and post-processing of data. This pattern is currently the most widely used, serving as a basis for building proofs of concept and seeking for quantum advantage in AI systems using gate-based quantum circuits.

\noindent \textit{Benefits:}
\begin{itemize}
    \item \textit{Efficiency.} The monolithic architecture typically includes only one encoding layer, or several in cases involving data re-uploading (see Figure~\ref{fig:cirquit}a and b), alongside a single measurement layer. Avoiding repeated time-consuming encoding and measurement procedures, as well as frequent communication between classical and quantum AI components -- which is typically managed via a network -- reduces both inference and training times.
    
    \item \textit{Simplicity and deployability.} A small number of components and classical communication channels between them enhances the simplicity and deployability of the systems.
\end{itemize}

\noindent \textit{Drawbacks:} 
\begin{itemize}
    \item \textit{Restricted scalability for NISQ computers.} For NISQ devices, directly encoding raw data into quantum circuits limits the volume of data that can be processed due to the restricted number of qubits available.  Additionally, NISQ-era computers impose restrictions on quantum circuit depth (defined as the maximum number of quantum gates applied sequentially) for monolithic architectures due to inherent noise that limits coherence times.  Note that many existing quantum encoding algorithms require a number of gates that scales exponentially with the number of input qubits, $n$ \cite[]{larose2020robust, PhysRevA.109.052423}. This restriction can be somewhat relaxed to O$(\text{poly}(n))$ using the approximate amplitude encoding \cite[]{PhysRevResearch.4.023136, PhysRevA.109.052423}.

    \item \textit{Restricted trainability at large scale.} The restricted trainability of quantum circuits for machine learning in the monolith architecture is related to the barren plateau issue [S99]. This phenomenon involves the exponential diminishing of the gradient of the cost function used to train quantum neural networks as the size of a quantum monolith system increases, impeding the training of quantum AI systems. 
\end{itemize}

\noindent \textit{Known uses:} 
\begin{itemize}
    \item \cite{Grant2018} [S27] utilised quantum circuits to perform binary classification of classical and quantum data, achieving superior accuracy by leveraging more expressive circuits and robustness to noise.
    \item \cite{PhysRevX.12.021037} [S22] incorporated quantum correlations into generative models, i.e., Bayesian networks, for unsupervised learning tasks.
    \item \cite{LVMXUC3B} [S72] proposed an experimental implementation of a simple quantum monolith framework for quantum generative adversarial networks using a superconducting processor with multiple qubits.
\end{itemize}

\noindent \textbf{\textit{SP-2: Multi-layer.}}

\begin{figure}[h!]
    \centering
    \includegraphics[height=0.18\linewidth]{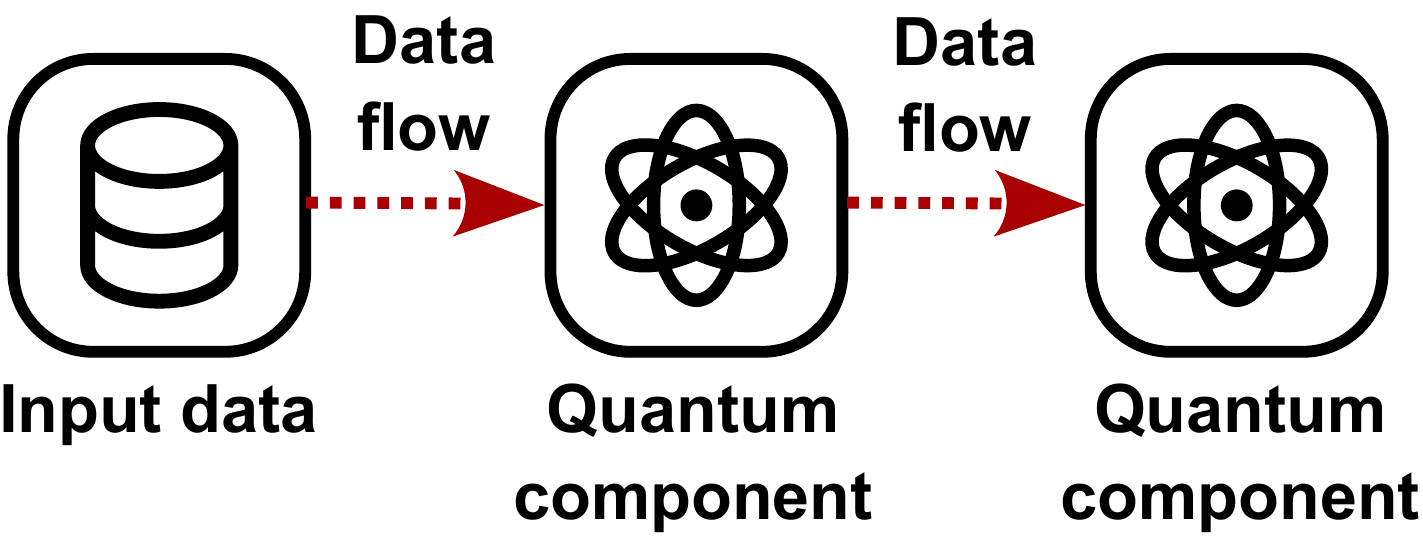}
    \caption{Multi-layer}
    \label{fig:multilayer}
\end{figure}

\noindent \textit{Summary:}
Incorporating multiple quantum components with learnable parameters, interconnected through classical channels, can significantly enhance the overall performance of an AI system by improving expressiveness and accuracy. Figure~\ref{fig:multilayer} illustrates a simple graphical representation of the multi-layer pattern.

\noindent \textit{Problem:} The problem to solve is enhancing expressivity of the quantum-enhanced model and increasing the number of trainable parameters within the constraints imposed by NISQ-era quantum computers, which limit the number of qubits and the depth of quantum circuits.

\noindent \textit{Solution:} Given the constraints of NISQ computers, the expressivity of the model and inference accuracy can be enhanced by integrating several quantum components, which are interconnected by classical channels. This setup functions similarly to how additional layers of neurons in deep artificial neural networks improve performance by allowing for more complex and hierarchical feature extraction. In particular implementations of this pattern, multiple quantum components can function as individual quantum neurons [S6], layers of neurons [S17, S35, S73, S110], or they can perform various tasks in the inference pipeline, such as feature extraction and subsequent feature classification in quantum support vector machines [S111].

\noindent \textit{Benefits:} 
\begin{itemize}
    \item \textit{Composability.} This architecture is utilised in quantum-classical hybrid deep neural networks where one quantum component is often considered as a so-called quantum neuron 
    [S6, S35, S73, S110]. Quantum neurons, connected through classical communication channels, can be arranged into neural networks in various configurations, enabling a wide range of systems with diverse parameters.
    \item \textit{Scalability for NISQ computers.} The quantum multi-layer pattern aims to address the limitations of NISQ-era quantum computers, which are constrained by the depth of quantum circuits and the number of qubits. A single neuron or perceptron, representing one layer in the architecture, can be implemented using only a few qubits and a limited number of entanglement gates, making it compatible with most existing quantum computers [S6].
    \item \textit{Trainability and expressivity.} The multi-layer architecture is capable of reproducing a probability distribution with a much steeper gradient near the classification line which was demonstrated by \cite{86SK6A24} [S35]. The quantum multi-layer architecture is also less susceptible to the barren plateau problem when many shallow quantum circuits are used.
\end{itemize}

\noindent \textit{Drawbacks:}
\begin{itemize}
\item \textit{Efficiency.} Introducing additional quantum components that exchange classical information requires extra encoding of classical information into quantum states. This encoding is usually characterised by exponential gate complexity and can pose a bottleneck for quantum speedup if quantum components are large-scale. Additionally, depending on the required accuracy, performing multiple measurements for each layer can become an efficiency bottleneck, especially during the training phase, where repeated gradient calculations for each neuron are performed.
\item \textit{Portability and deployability.} This architecture requires low-latency, high-speed communication channels between quantum components and the storage of intermediate data transferred through those channels. Therefore, it demands specialised hardware infrastructure, which is not always provided by default in quantum computers.

\end{itemize}
\noindent \textit{Known uses:} 

\begin{itemize}
    \item \textit{Classification.} Several quantum layers interconnected by classical communication channels has been utilised in deep neural networks for classification problems
    [S17; S35; S73].
    \item \textit{Generalisation from fewer examples.} This pattern can help create neural networks that generalise better from fewer examples, which is particularly beneficial for applications with limited data [S6].
    \item \textit{PennyLane.} PennyLane\footnote{\url{https://pennylane.ai/}} is a software framework that provides differentiable quantum nodes with classical data inputs and outputs. These nodes can be sequentially stacked or used in conjunction with classical components.
\end{itemize}

\noindent \textbf{\textit{SP-3: Quantum Feature Engineering}}

\begin{figure}[!h]
    \centering
    \includegraphics[height=0.18\linewidth]{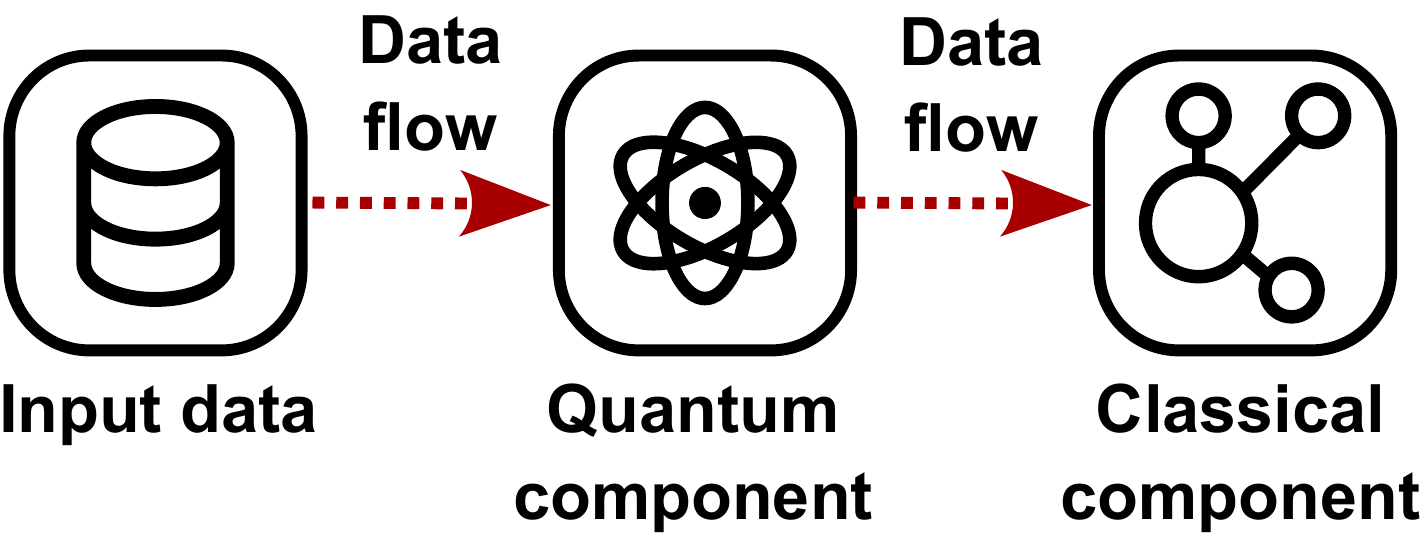}
    \caption{Quantum feature engineering}
    \label{fig:feature}
\end{figure}

\noindent \textit{Summary:} 
This pattern applies when feature engineering is delegated to a quantum computer, which serves as the first step in the inference pipeline by operating directly on the input data. The classical inference engine then carries out subsequent steps in the pipeline.
Figure~\ref{fig:feature} illustrates a simple graphical representation of the quantum-feature-engineering pattern. 

\noindent \textit{Problem:} Quantum computing offers benefits in the feature extraction procedure. However, limitations of NISQ-era quantum computers, such as the number of qubits and the depths of quantum circuits, prevent the efficient use of quantum components for further inference tasks.

\noindent \textit{Solution:} 
This pattern leverages the power of quantum computing to initially extract features from raw data, taking advantage of its computational strengths and large dimensionality of the underlying Hilbert space [S75]. These extracted features are subsequently transferred to a classical system, which performs further analysis and inference using conventional machine learning algorithms. This pattern is often employed in support vector machines, where the quantum component is responsible for evaluating the kernel function. \cite{PhysRevLett.122.040504} [S75] distinguishes between two types of quantum kernel models: the implicit approach, where the quantum device evaluates only the kernel function, and the explicit approach, where both kernel evaluation and classification are handled by the quantum computer. This pattern describes the former case.

\noindent \textit{Benefits:} 
\begin{itemize}
    \item \textit{Efficiency.} The quantum-feature engineering can help extract high-quality features, enhancing the performance of machine learning tasks. Quantum feature maps can, in principle, be computed with exponential speed-up for certain well-suited problems [S3]. For more practical applications, reports indicate that quantum support vector machines may operate with sub-quadratic run-time complexity [S111].
    \item \textit{Robustness.} Several publications highlight the robustness of quantum feature engineering against noise in both data and hardware [S3; S46; S93; S102].
    
\end{itemize}

\noindent \textit{Drawbacks:} 
\begin{itemize}
    \item \textit{Restricted scalability for NISQ computers.} Currently, quantum feature engineering can only be directly applied to data with relatively small dimensions due to the limited number of qubits available in NISQ-era quantum computers.
\end{itemize}

\noindent \textit{Known uses:} 

\begin{itemize}
    \item \cite{EBKMYZ93} [S7] proposed the utilisation of quantum feature engineering to classify brain neuron morphologies using multiclass classification with quantum kernel methods. They examined the impact of feature engineering on classification accuracy and found that quantum kernel methods performed similarly to classical methods.

    \item \cite{6ANRF8NT} [S13] utilised a quantum distance classifier and a simplified quantum-kernel support vector machine on the 15-qubit IBM Melbourne quantum computer to address a classification problem using real clinical datasets.
    
    \item \cite{HRPF3VDM} [S87] employed quantum computing techniques for feature extraction in image scene classification.
\end{itemize}

\noindent \textbf{\textit{SP-4: Quantum head}}

\begin{figure}[!h]
    \centering
    \includegraphics[height=0.33\linewidth]{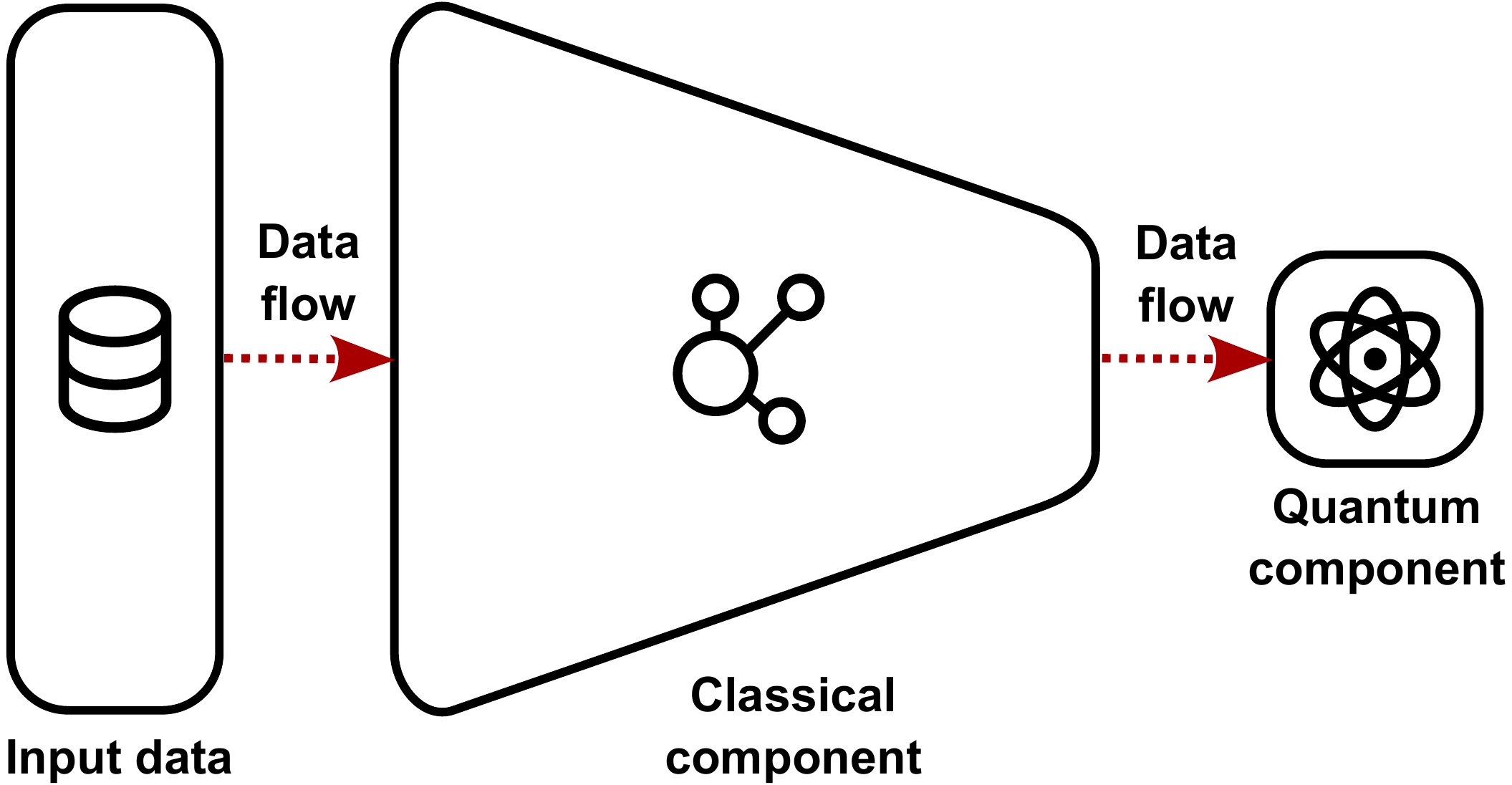}
    \caption{Quantum head}
    \label{fig:head}
\end{figure}

\noindent \textit{Summary:} This pattern implements an architecture where a quantum inference engine operates in the pipeline after a classical neural network. The classical component is employed to reduce the dimensionality of the input data. Figure~\ref{fig:head} illustrates a simple graphical representation of the quantum head pattern.

\noindent \textit{Problem:} In real-world applications, input data often has large dimensions and cannot fit into modern NISQ computers. This limitation poses challenges in processing large-dimensional input data.

\noindent \textit{Solution:} To leverage the quantum advantage with high-dimensional graphical data, a hybrid architecture has been proposed. In this architecture, the dimensionality of the input data is first reduced through data pre-processing and feature extraction on classical computers. Subsequent inference is then performed on a quantum computer. This pattern works well with pre-trained deep neural networks taking advantage of the quantum transfer learning technology 
[S11-12; S18; S21; S33; S84; S111].

\noindent \textit{Benefits:} 
\begin{itemize}
    \item \textit{Scalability for NISQ computers.} Implementations of this pattern require quantum components with a small number of qubits and shallow quantum circuits, making the system compatible with most existing NISQ computers. This pattern makes it possible to use NISQ-era quantum computers for practical applications that process high-dimensional data.
    \item \textit{Efficiency.} This architecture aims to substitute the several last layers in a classical neural network, most often the last fully-connected layers, by a quantum neural network, improving performance and reducing number of parameters 
    [S1; S11-12; S18; S21; S31; S33; S38; S48; S52; S54; S67-68; S78; S84; S93-94;
S107; S111].

    \item \textit{Trainability with transfer learning.} For gradient descent methods used in training neural networks, computing gradients for quantum components and transferring this information between classical and quantum elements presents a challenge. To overcome these difficulties, quantum transfer learning technology can be utilised with this pattern.
\end{itemize}

\noindent \textit{Drawbacks:} 
\begin{itemize}
\item \textit{Portability and deployability.} The Quantum Head pattern implies broadband and high-speed communication channels between classical and quantum components, which is not always provided by default.

\end{itemize}

\noindent \textit{Known uses:} 
\begin{itemize}
    \item \cite{YMMEXC29} [S52] and  \cite{KMRPKRUT} [S93] tested a hybrid architecture, which consists of an input layer, several classical neural network hidden layers, and one quantum layer. In these works, the authors applied this architecture to supply chain backorder prediction and abstract binary classification, respectively. To compute the gradients of quantum layers, the authors in~\cite[]{pennylane2022} utilised PennyLane, which introduced the concept of differentiable quantum nodes that can be used in conjunction with classical layers.
    \item \cite{GSDMNCPI} [S11] utilised the quantum head pattern with transfer learning for classifying the MNIST dataset.
    \item \cite{QV4N5TVV} [S68] employed the quantum head pattern and a tree tensor network as part of reinforcement learning algorithms for AlphaZero.
\end{itemize}

\noindent \textbf{\textit{SP-5: Quanvolution}}

\begin{figure}[!h]
    \centering
    \includegraphics[height=0.33\linewidth]{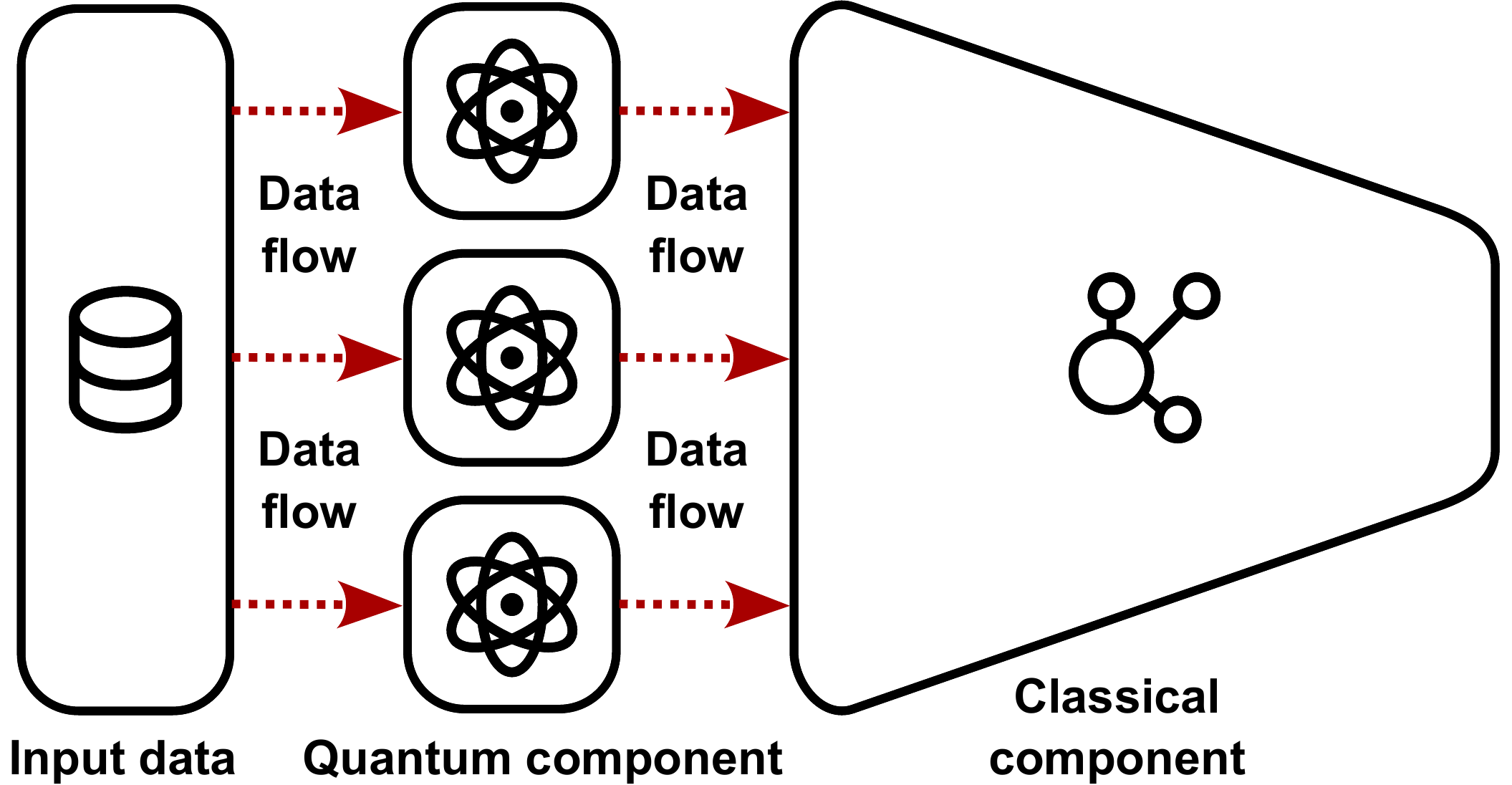}
    \caption{Quanvolution pattern}
    \label{fig:quanvolution}
\end{figure}

\noindent \textit{Summary:} This architectural solution replaces the first several layers in classical convolutional neural networks by parameterised quantum circuits operating in parallel or sequentially. The circuit has a relatively small number of qubits and operates as a convolutional filter. Figure~\ref{fig:quanvolution} illustrates a simple graphical representation of the quanvolution pattern.

\noindent \textit{Problem:} The challenge is to adapt the quantum feature extraction pattern (as detailed in the previous pattern) for high-dimensional data, given the constraints of NISQ-era quantum computers.

\noindent \textit{Solution:} A small quantum circuit with trainable parameters is used as a convolutional filter, replacing the first several layers of a convolutional neural network. The quantum circuit sequentially scans over the input data tensor, processing a small data window at a time.

\noindent \textit{Benefits:}
\begin{itemize}
    \item \textit{Scalability for NISQ computers.}  Regardless of the dimensionality of the input data, implementations of this pattern require shallow quantum circuits with a small number of qubits, making the system compatible with most existing NISQ computers. This pattern enables the system to process high-resolution images on NISQ devices [S23].
    \item \textit{Applicable to spatial multi-dimensional data} The quanvolutional filter processes data locally while preserving information on spatial order and dimensionality, even when quantum circuits with 1D input are used [S10, S14].
    \item \textit{Reduced number of parameters and space complexity.} Replacing the first several layers of a convolutional neural network by quantum circuits reduces the number of parameters without compromising the expressibility of the model [S10; S23; S26].
    
    \item \textit{Distributability} This pattern allows for parallelization, as the quanvolution filters can operate simultaneously, share the same parameters, and are interconnected via classical channels.
    
    \item \textit{Trainability} There were no indications of the Barren Plateau issue for this pattern, as the quantum circuits used are mostly shallow and have few trainable parameters.
\end{itemize}

\noindent \textit{Drawbacks:} 
\begin{itemize}
    \item \textit{Portability and deployability} This pattern requires frequent data exchange transactions between classical and quantum components. As a result, the system can only be deployed on computers where quantum processing units, CPUs and/or GPU are closely integrated. If the components are designed to exchange data over a network, a high-bandwidth, fast connection must be established. Therefore, a highly specialised hardware environment must be established for this pattern to function effectively.
    \item \textit{Efficiency} Using a quantum circuit as a convolutional filter requires multiple encodings and measurements, which can become a performance bottleneck.
\end{itemize}

\noindent \textit{Known uses:}

\begin{itemize} 
    \item \cite{FKTMJWRS} [S90] have applied the quanvolutional pattern  to perform multi-class classification for cardiovascular diseases.
    \item \cite{4MJ6LUYI} [S10] utilised the quanvolutional pattern as a part of a quantum-classical convolutional neural network developed for predicting binding affinity in drug design.
    \item \cite{5X7N9PDN} [S97] for point cloud data processing in classification applications.
\end{itemize}

\noindent \textbf{\textit{SP-6: Intermediate Quantum Layer}}

\begin{figure}[!h]
    \centering
    \subfigure[]{\includegraphics[height=0.33\linewidth]{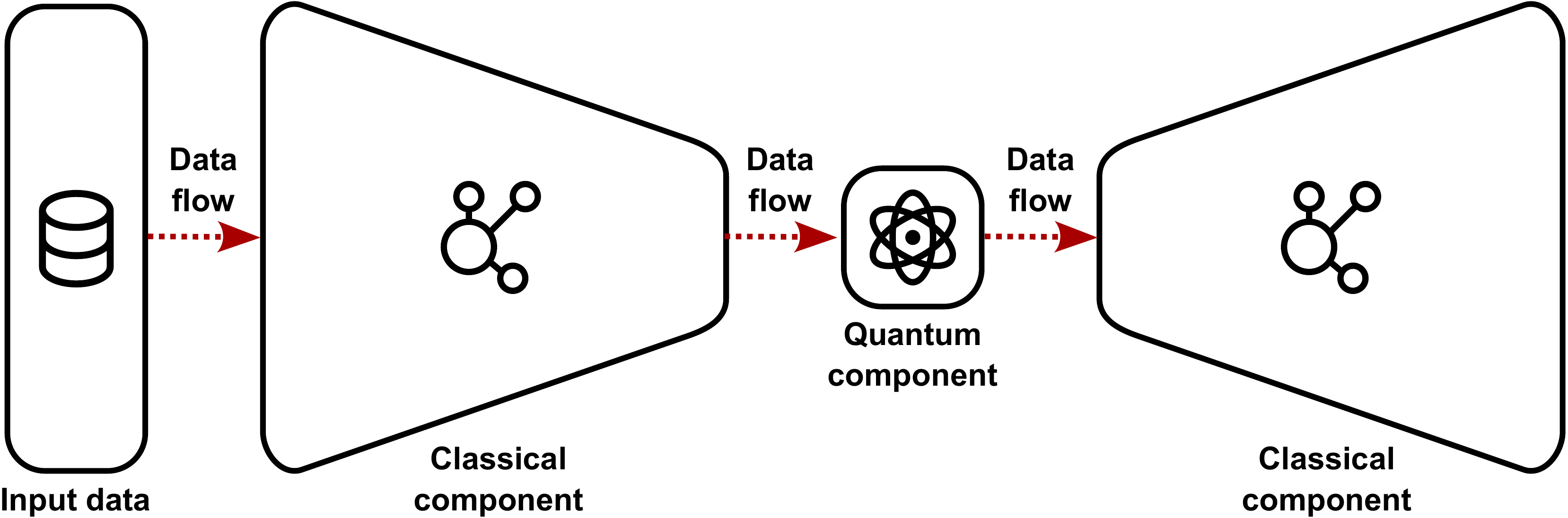}}
    \subfigure[]{\includegraphics[height=0.33\linewidth]{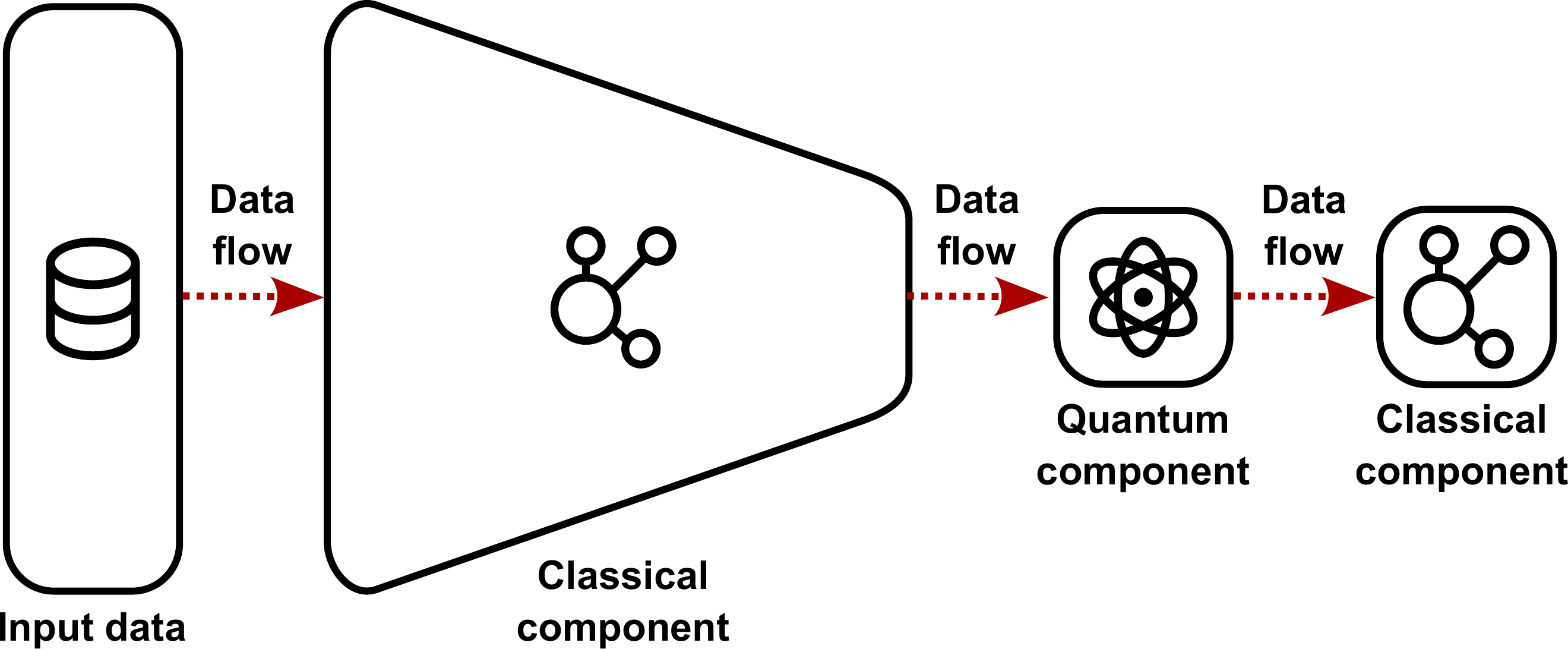}}
    \caption{Two possible implementations of the Intermediate Quantum Layer pattern with a) low-dimensional and b) high-dimensional output.}
    \label{fig:inter}
\end{figure}

\noindent \textit{Summary:} Similar to the Quantum Head pattern, this pattern involves a pipeline of classical and quantum inference engines. In this approach, classical components are employed in both the initial and final stages of the inference pipeline, while the quantum component is used in the intermediate stage. Figure~\ref{fig:inter} illustrates a simple graphical representation of the intermediate quantum layer pattern in two configurations, where the final classical component produces either low-dimensional or high-dimensional output.

\noindent \textit{Problem:} The challenge lies in achieving and maximising quantum advantage for AI systems within the constraints of NISQ-era quantum computers, which include a limited number of qubits and restricted circuit depth, while handling high-dimensional input and/or output data.

\noindent \textit{Solution:} The limitations of NISQ-era quantum hardware, particularly the restricted depth of quantum circuits, can be mitigated by embedding a quantum inference engine into a pipeline with classical inference engines, replacing several intermediate layers of the inference pipeline. In this approach, the classical components reduce diminsionality input data, performing initial stages of the feature extraction, and also processes data output from the quantum component interpreting results of the quantum state measurements. This solution can be viewed as a combination of the Quantum Head pattern and Quantum Feature Engineering where classical components participate both in quantum embedding for feature extraction and feature processing. This pattern is often utilised in applications addressing classification problems. In a classifier, classical components reduce the dimensionality of the input data through feature extraction and interpret the output from the quantum component 
[S4; S16; S24; S29; S33-34; S36; S39; S49; S95; S112]. Additionally, the quantum intermediate layer can serves as a blueprint for systems implementing quantum variational autoencoders 
[S28, S30].

\noindent \textit{Benefits:} 
\begin{itemize}

\item \textit{Scalability for NISQ computers.} Scalability for both input and output data is ensured by the classical components, which adjust the data to the available number of qubits.
 
\item \textit{Reduced number of parameters and space complexity.} Replacing several layers of a classical neural network by quantum circuits reduces the number of parameters without compromising the expressibility of the model [S16; S28].

\end{itemize}

\noindent \textit{Drawbacks:} 
\begin{itemize}
\item \textit{Portability and deployability.} The Intermediate Quantum Layer pattern implies high-speed communication channels between classical and quantum components, and requires significant computational resources for classical components, which are not always available by default.
\end{itemize}

\noindent \textit{Known uses:} 

\begin{itemize}
    \item \cite{SP42M5A9} [S39] used the intermediate quantum layer to improve financial forecasting.
    \item \cite{TUNHBV9C} [S49] and \cite{C75NJPQ6} [S33] applied this pattern in a hybrid architecture for multiclass classification of remote sensing imagery and classification of computed tomography scans of lungs to detect COVID-19, respectively.
    \item \cite{Q5GLRGG8} [S30] used the intermediate quantum layer in the architecture of an  autoencoder for anomaly detection in a dataset.
\end{itemize}

\noindent \textbf{\textit{SP-7: Quantum accelerator}}

\begin{figure}[!h]
    \centering
    \includegraphics[height=0.5\linewidth]{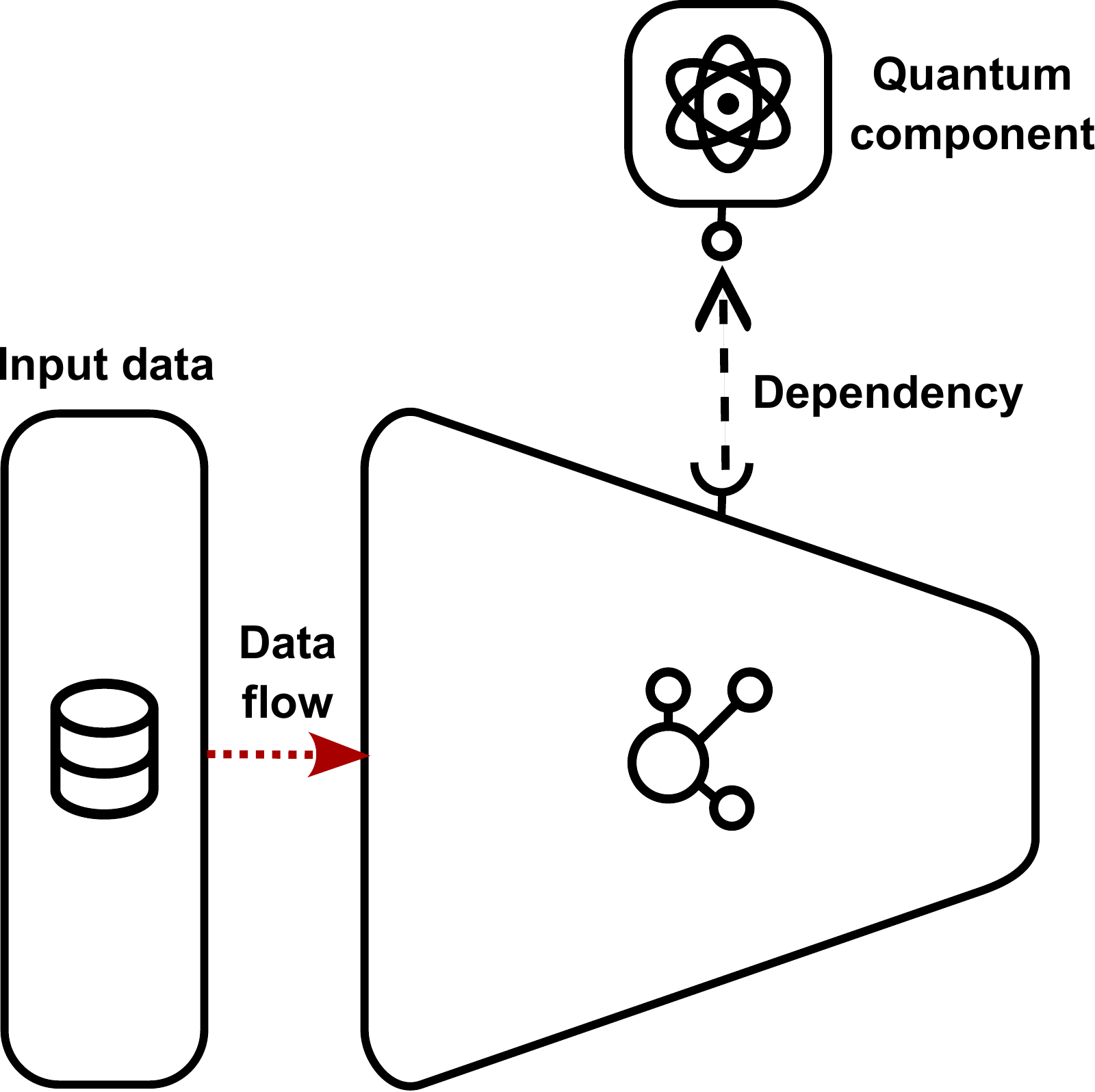}
    \caption{Quantum accelerator}
    \label{fig:accelerator}
\end{figure}

\noindent \textit{Summary:}  This pattern employs a quantum component to evaluate a specific, well-defined function within the system. This quantum component, called a quantum accelerator, usually has a classical analog with lower performance, introduces dependencies on operations executed by a quantum computer, and typically does not possess trainable parameters. Figure~\ref{fig:accelerator} illustrates a simple graphical representation of the quantum accelerator pattern. 

\noindent \textit{Problem:} Quantum computing promises to offer an exponential advantage for certain tasks, such as linear algebra operations and optimization. The challenge lies in identifying practical methods to leverage this advantage to enhance the performance of AI systems. This challenge is related to the trade-off between integration costs, performance, deployability, and maintenance.

\noindent \textit{Solution:} Similarly to the acceleration of neural networks on classical computing platforms such as CPUs, GPUs, FPGAs, and ASICs, the quantum processing unit can be utilised to accelerate computations as well. At the software level, a quantum algorithm that provides a quantum advantage is wrapped in a function and integrated with a classical inference engine. This function is further mapped onto a service or microservice offered by a particular quantum hardware, which interacts with the system via API calls.

\noindent \textit{Benefits:} 
\begin{itemize}
    \item \textit{Efficiency.} It is anticipated that the quantum advantages provided by quantum algorithms can be translated into practical performance enhancements and improvements for AI systems.
\end{itemize}

\noindent \textit{Drawbacks:} 
\begin{itemize}
    \item \textit{High latency numbers.} Depending on the frequency of evaluating the accelerated function, this pattern may require a tight integration of classical and quantum components as well as intense data exchange between them. When accessible via a network, latency must be considered, and parallelization should be employed to ensure a reasonably fast system response [S41].
\end{itemize}

\noindent \textit{Known uses:} 
\begin{itemize}
    \item \cite{Cherrat2024quantumvision} [S85] utilize quantum circuits to speed up matrix multiplication routines that are utilised in vision transformer models, and \cite{4SCZCHNF} [S41] have developed an accelerator for a data-driven function.
    \item For classical neural networks, \cite{MZM6BYYV} [S106] have accelerated the training process, that is based on the stochastic gradient descent method. They implemented this by employing a variant of the Harrow-Hassidim-Lloyd algorithm, a highly efficient quantum algorithm for sparse matrix inversion. This algorithm solves the problem in $\log(n)$ time for suitably conditioned $n \times n$ sparse matrices. 
    \item \cite{NUEYEGIB} [S40] have improved the training speed for support vector machine algorithm applied to the fraud detection problem by using quantum annealing solvers. The authors have reformulated the problem of obtaining kernel functions for support vector machines as a quadratic unconstrained binary optimisation (QUBO) problem. Subsequently, the QUBO problem has been solved using quantum annealing solvers.
\end{itemize}

\subsection{Quantum middleware architectural patterns (MP)}
\label{sec:mp}

Quantum middleware architectural patterns are defined within the context of the interaction of quantum components with traditional classical software, operating systems, and network infrastructure. Many of these interactions imply the existence of a middleware software layer that can organise the communication, orchestration, and scheduling of quantum components and workflows. We have identified the following architectural patterns related to middleware in quantum AI systems:

\begin{table*}[!ht]
    \centering
        \caption{Quantum Middleware Architectural Patterns}
        \footnotesize
\begin{tabularx}{\textwidth} { 
  | >{\raggedright\arraybackslash}p{\dimexpr.15\linewidth-2\tabcolsep-1.3333\arrayrulewidth} 
  | >{\raggedright\arraybackslash}p{\dimexpr.7\linewidth-2\tabcolsep-1.3333\arrayrulewidth}
  | >{\raggedright\arraybackslash}p{\dimexpr.15\linewidth-2\tabcolsep-1.3333\arrayrulewidth} | }
 \hline
 Name of the pattern (ID)  & Summary & Refs. \\ 
 \hline Service or micro-service wrapper (MP-1) & 
In the context of quantum middleware, this pattern enables seamless communication and data exchange between classical and quantum components by wrapping quantum resources into classical services or small, loosely coupled, and independently deployable micro-services. & [S20; S44-45; S53; S55; S80-81; S83; S92; S96; S109] \\ 

 \hline Quantum API gateway (MP-2) & The Quantum API Gateway, being one of the middleware services, is responsible for routing API requests appropriately as well as identifying the most suitable quantum computer to execute a specific quantum service at runtime.  & [S80-81; S83; S104; S109]\\
 \hline 
 Quantum workflows orchestration (MP-3) & In the context of hybrid quantum-classical workflows or quantum micro-service architecture, this pattern is designed to automatically deploy quantum components to the most suitable quantum provider, as well as provide automated scheduling of quantum programs to run continuously, or according to time-based or event-based schedules. & [S2; S32; S42; S47; S82] \\
 \hline 
\end{tabularx}
    \label{tab:middleware}
\end{table*}

\noindent \textbf{\textit{MP-1: Service or micro-service wrapper}}

\noindent \textit{Summary:} In the context of quantum middleware, this pattern enables seamless communication and data exchange between classical and quantum components by wrapping quantum resources into classical services or small, loosely coupled, and independently deployable micro-services. 

\noindent \textit{Problem:} Since most existing quantum computers are neither portable nor mass-produced, access is provided via local and global networks by major players such as Amazon,\footnote{\url{https://aws.amazon.com/braket/}} IBM,\footnote{\url{https://quantum.ibm.com/}} Google,\footnote{\url{https://quantumai.google/}} D-Wave,\footnote{\url{https://www.dwavesys.com/}} Microsoft,\footnote{\url{https://quantum.microsoft.com/}} and Rigetti.\footnote{\url{https://www.rigetti.com/}} Quantum Brilliance is the only known startup developing a portable quantum computer, making it a notable exception.\footnote{\url{https://quantumbrilliance.com/}} The existing hardware API provided by the quantum services vendor are typically characterised by technology and vendor lock-in, may offer limited capabilities, and often do not align with application requirements. This presents challenges for the accessibility, deployability, flexibility, resilience, and maintainability of quantum-enhanced AI systems. 

\noindent \textit{Solution:} The solution is to wrap the functionalities provided by the quantum hardware within a service facade that interacts with the provided quantum resources and exposes a required service interface to the consumers. This facade is typically built on top of the cloud computing infrastructure provided by vendors through models such as Quantum Computing as a Service (QCaaS) model 
[S53, S55, S92] or Quantum Infrastructure as a Service (QIaaS)  model 
[S80] or Platform as a Service (QPaaS)\footnote{\url{https://www.ibm.com/quantum/blog/quantum-reliability}}. These models allows developers for accessing high-level abstractions for quantum hardware functionalities over the cloud, and has already become a standard cloud quantum model offered by industry vendors such as IBM, Microsoft Azure\footnote{\url{https://azure.microsoft.com/en-us/products/quantum}}, and Amazon Braket. A comprehensive review of cloud quantum computing can be found in \cite[]{nguyen2024}. 

Note that the quantum software full-stack often follows a layered architecture, as presented in [S44]. When applied within the context of a service-oriented architecture, this layered structure gives rise to a service layer pattern. In quantum systems, the service layer functions as a wrapper for higher-level abstractions rather than organising business logic, making it functionally equivalent to the service wrapper pattern.

\noindent \textit{Benefits:} 
\begin{itemize}
    \item \textit{Accessibility.} The micro-services wrapper pattern increases accessibility by decoupling quantum software from the constraints of platform-specific programming languages, hardware, and associated services. This abstraction facilitates broader interoperability and flexibility, enabling quantum applications to seamlessly integrate across diverse technological environments 
    [S81].
    \item \textit{Scalability and composability.} This pattern enhances scalability and composability by leveraging specialised, unified, and standardised API functions 
    [S20; S45; S83]. 
    \item \textit{Reliability.} This pattern optimises the interface between client interactions and quantum backend services, enhancing reliability and ensuring stable and consistent performance across different operational scenarios 
    [S55].
    \item \textit{Deployability.} Wrapping quantum functionalities into micro-services enhances deployability and maintainability of the quantum AI applications [S53; S47; S55; S81; S83; S109].
\end{itemize}

\noindent \textit{Drawbacks:} 
\begin{itemize}
    \item \textit{Complexity.} Wrappers add complexity. Creating high-level abstractions by aggregating and wrapping provided service interfaces can introduce additional bugs and errors in the code.
\end{itemize}

\noindent \textit{Known uses:} 
\begin{itemize}
    \item \cite{XUULPYC3} [S80] used Amazon Braket to deploy quantum services by wrapping them in a classical service. They used prime factorisation and the traveling salesman problem as examples to study the effect of different quantum hardware on the service performance.

    \item Several authors such as \cite{10.1007/978-3-030-72369-9_2} [S92], \cite{IVYVVBCE} [S109] and \cite{SRBWSLQR} [S83]  propose a service model called Quantum Application as a Service (QAaaS) or Quantum Software as a Service (QSaaS). This model provides users access to an extensive array of preconfigured quantum software tools and applications and encapsulates the most common quantum applications and their deployment logic, including the quantum circuit along with data pre- and post-processing. It also provides an API that can be called within other applications.

    \item \cite{3V7TMZQ4} [S53] proposed a holistic Quantum Function-as-a-Service framework (QFaaS), which leverages the advantages of the serverless model.

\end{itemize}

\noindent \textbf{\textit{MP-2: Quantum API gateway}}

\noindent \textit{Summary:} The Quantum API Gateway
, being one of the middleware services, is responsible for routing API requests appropriately 
[S104] as well as identifying the most suitable quantum computer to execute a specific quantum service at runtime.
 
\noindent \textit{Problem:} The problem is establishing a way for clients of a quantum microservices-based application to access individual quantum services and determine the most suitable host for running quantum algorithms, all while keeping the application hardware-agnostic.

\noindent \textit{Solution:} The Quantum API Gateway serves as the system's entry point, routing requests to the most appropriate quantum microservices. It also aggregates results to return to the consumer or creates a fine-grained API tailored to each specific client type. In making these decisions, the Quantum API Gateway utilizes various available information, such as the availability of quantum computers, economic costs associated with hardware usage, and estimated response times. It also considers the current characteristics of quantum computers, including qubit topology, error rates, and fidelity 
[S83]. This gateway is an adaptation of the API Gateway pattern from traditional microservice architecture, tailored to address the unique deployment requirements of quantum services \cite[]{SRBWSLQR} [S83]. 

\noindent \textit{Benefits:} 
\begin{itemize}
    \item \textit{Deployablity.} The Quantum API Gateway optimises the deployment strategy by determining, at runtime, which available quantum computer is best suited for executing a specific quantum service. This approach enhances the efficiency of the quantum service invocation process 
    [S83].
    \item \textit{Hardware decoupling.}  The use of this pattern also facilitates decoupling between services and between services and hardware, as the API Gateway conceals this complexity from service clients 
    [S80-81].
    \item \textit{Services composability.} This approach facilitates the composability of heterogeneous microservices by breaking down a single invocation into calls to various services without their awareness.
\end{itemize}

\noindent \textit{Drawbacks:}
\begin{itemize}
    \item \textit{Performance.} A quantum API Gateway can introduce an additional single point of failure and increase response time. If not scaled properly, it can become a performance bottleneck. 
\end{itemize}
\noindent \textit{Known uses:} 
\begin{itemize}
    \item As a proof of concept, the quantum API gateway has been implemented for the Amazon Braket quantum computing platform \cite[]{SRBWSLQR} [S83]. 
    \item Authors of several works \cite[]{XUULPYC3, Y97WFA6Q, IVYVVBCE} [S80-81; S109] consider the quantum API gateway to be a necessary integral component of the quantum service-oriented architecture.
    \item \cite[]{PQR7YDXZ} [S104] propose a hybrid cloud-based reference architecture for the quantum-science gateway that reduces the complexity of using quantum computing resources and increases their accessibility.
\end{itemize}

\noindent \textbf{\textit{MP-3: Quantum workflows orchestrator}}

\noindent \textit{Summary:} In the context of hybrid quantum-classical workflows~\cite[]{WYK2PHBD, VDA7XFCL, leymann2021hybrid} or quantum micro-service architecture~\cite[]{D9TYIEZ6}, this pattern is designed to automatically deploy quantum components to the most suitable quantum provider [S32, S82], as well as provide automated scheduling of quantum programs to run continuously, or according to time-based or event-based schedules [S109].

\noindent \textit{Problem:} In quantum AI systems, the technological stack significantly expands in both size and complexity, leading to increased interdependencies and interactions among various system components. The challenge involves identifying, deploying, and executing quantum components while ensuring compatibility, performance, and availability of the entire application, given that quantum hardware platforms have NISQ-era limitations.

\noindent \textit{Solution:} To address these challenges, the software development and architectural designs of such systems necessitate tools that can structure software transactions into workflows. A workflow is defined as a technology that specifies the partial order of a collection of heterogeneous activities aimed at achieving a composite goal~\cite[]{leymann2021hybrid}. The specific sequence of these activities is dictated by a workflow model. Typically represented as nodes in a directed graph, workflow models delineate the required tasks, their execution sequence, and the data flow between them. A workflow engine, or orchestrator, facilitates the automatic orchestration, deployment, and execution~\cite[]{WYK2PHBD}. The orchestrator is often implemented as middleware that analyses high-level programming language code, identifies quantum components, executes a classic-quantum split, and generates corresponding API functions. This middleware pattern, which can be implemented on either the server side or client side, generally includes components such as an orchestrator~\cite[]{IVYVVBCE, D9TYIEZ6} [S109, S82], a language translator/generator~\cite[]{IVYVVBCE, SN5TJDV6}, and a scheduler~\cite[]{IVYVVBCE, D9TYIEZ6}.

\noindent \textit{Benefits:} 
\begin{itemize}
    \item \textit{Autonomy.} The scheduling and deployment of quantum workflows are automated, requiring no human intervention.
    \item \textit{Hardware decoupling.} This pattern promotes the separation of quantum software from specific hardware and interface constraints.
     \item \textit{Availability.} The primary purpose of this pattern is to ensure the availability of quantum services for the application when needed.
\end{itemize}

\noindent \textit{Drawbacks:} 
\begin{itemize}
    \item \textit{Reliability.} This pattern introduces a single point of failure, which means that if one part of the middleware fails, it could potentially disrupt the entire service, causing system-wide outages.
    \item \textit{Response time.} This pattern can result in extended response times in the development of quantum AI systems. Each added middleware layer introduces additional complexity, potentially delaying the delivery of data and responses to user requests.
    \item \textit{Performance.} As increasing volumes of processes and data are channeled through the additional middleware layers inherent in this pattern, these layers may become overwhelmed. This congestion can slow down throughput and consequently degrade the overall performance of the system.
\end{itemize}

\noindent \textit{Known uses:} 
\begin{itemize}
    \item \cite{10234288} [S2] considered integration patterns in quantum middleware and used well-established high-performance computing abstractions for managing workloads, tasks, and resources to integrate quantum computing into HPC systems.
    \item \cite{9590459} [S32] presented a prototypical implementation that integrates topology and workflow orchestrations, utilising the Camunda workflow engine and the OpenTOSCA ecosystem.
    \item \cite{MWWGF84E} [S47] propose an on-the-fly cost minimiser for quantum microservices, as part of a middleware service. This tool identifies the most cost-effective classical and quantum machines for deploying hybrid microservice-based applications. 
    
\end{itemize}

\section{Rationale for using quantum computing in AI systems (RQ2)}
\label{sec:advantage}

From a theoretical standpoint, quantum technology offers the potential to enhance the performance of current AI systems. During our systematic mapping study, we observed preliminary attempts to realise this potential using existing NISQ-era quantum computers. The outcomes of these efforts are compiled and summarized in Table~\ref{tab1} and detailed below.

\begin{table}[t!]
    \centering
        \caption{Reported reasons of utilising quantum computing in AI systems}
        \footnotesize
\begin{tabularx}{\linewidth} { 
  | >{\raggedright\arraybackslash}p{\dimexpr.33\linewidth-2\tabcolsep-1.3333\arrayrulewidth} 
  | >{\raggedright\arraybackslash}p{\dimexpr.67\linewidth-2\tabcolsep-1.3333\arrayrulewidth} | }
 \hline
 Reported advantage & Refs. \\ 
    \hline
   Comparable or slightly better inference accuracy & [S1; S11; S18-19; S24; S26; S30-31; S34-35; S36; S38; S49; S52; S54; S57; S62; S67; S76; S79; S87; S94; S96-97]\\ 
   \hline
  Training speed-up & [S5; S40; S94; S103; S106; S110] \\ 
    \hline
  Inference speed-up  &  [S3; S26; S36; S56; S77; S98; S111; S113]\\
    \hline
  Reduction number of model parameters & [S10-11; S16; S23; S26; S37-38; S63; S65; S87; S112]
  \\
    \hline
  Storage & [S6; S26; S28; S37; S73; S112] \\
    \hline
  Sampling advantage & [S22; S71; S100] \\
    \hline
 Adversarial &  [S4; S8; S25; S29; S58; S107] \\
   \hline
 Security & [S86] \\
   \hline
 Noise robustness &  [S4; S79; S93; S102] \\
   \hline
 Need of randomness & [S101] \\
  \hline 
\end{tabularx}
    \label{tab1}
\end{table}

\noindent \textbf{Accuracy of quantum machine learning.} One of the findings from our review is that current research does not provide evidence that QML can systematically outperform classical state-of-the-art algorithms in any practically relevant application in terms of accuracy. However, it is notable that NISQ-era QML is often capable of achieving accuracy comparable to many classical algorithms for small input data tensors. The papers selected in this study indicate comparable or in rare cases slightly enhanced inference accuracy across all the identified quantum-classical split architectures. Specifically, such accuracy is reported for the monolith architecture [S79; S96]
, multi-layer architecture [S35]
, feature extraction [S19; S76; S87]
, quantum head [S1; S11; S18; S31; S38; S52; S54; S67; S94]
, quanvolution [S26; S62; S97]
, intermediate quantum layer [S24; S30; S34; S36; S49]
, and quantum accelerator with a quantum annealer [S40; S57]
.

Most of these results have been derived from numerical experiments conducted on classical computers, under the assumption of fault-tolerant quantum computers. In 
 [S24], the authors incorporate a noise model in their simulations to replicate the noise characteristics of NISQ devices. Several attempts to run algorithms on actual IBM NISQ architectures and compare them with simulations indicate a decrease in accuracy due to noise [S35; S76; S87]
 . Additionally, an attempt to run a quantum neural network on a real quantum computer, such as Rigetti, for a classifier-based model for botnet attack detection, achieved accuracy comparable to existing state-of-the-art systems 
  [S96].

\noindent \textbf{Inference speed-up.} Many papers report the potential for algorithmic inference speed-up in quantum AI through the use of quantum neural networks 
 [S26; S77; S113; S125] and quantum kernels for support vector machines 
[S3; S98; S111]. This speed-up is typically associated with the reduced time complexity reported for quantum algorithms compared to their classical analogs 
[S26; S77; S98; S111; S113]. Due to the limited operational speed of current quantum gates, achieving actual performance gains on quantum hardware from the speedup associated with reduced time complexity in algorithms may require impractically large problem sizes. Several numerical experiments on classical simulations of quantum computers confirms faster execution times compared to classical algorithms for quantum neural networks 
[S36] and quantum kernels 
[S56]. The fact that speed-up can be achieved even in simulations lays the foundation for quantum-inspired machine learning systems. Quantum-enhanced kernels for support vector machines have been also successfully run on an IBM quantum computer, although their execution times were not reported 
[S98; S111]. 

One of the key challenges in achieving practical speedups in QML is the time/gate complexity of encoding classical information into quantum states. A potential solution to this challenge is the use of Quantum Random Access Memory \cite[]{QRAM}, which can reduce this complexity and improve data access speed. However, quantum memories require fault-tolerant quantum computers and remain an active area of research, with practical, large-scale implementations still yet to be realised. Another approach aims to leverage approximate encoding methods and incorporates variational, genetic, and matrix product state algorithms \cite[]{Usman2024}.

\noindent \textbf{Reduce model parameters.} Another advantage that gate-based quantum computing systems offer to AI is a substantial reduction in the number of model parameters without loss of performance 
[S10-11; S16; S23; S26; S37-38; S63; S65; S87; S112]. It is anticipated that quantum models with fewer parameters can achieve the same performance as classical models. 

\noindent \textbf{Reduce space complexity.} Quantum computing also offers advantages in information storage through the reduced space complexity of certain quantum algorithms 
[S26; S28; S125] and the informational capacity of qubits 
[S6; S73]. Specifically, $n$ qubits can represent a data point with
$M=2^n$ features using the superdense coding \cite[]{larose2020robust}.

\noindent \textbf{Accelerating training of AI systems.} Quantum technology has the potential to accelerate the training process of classical AI systems. One approach is to exploit adiabatic quantum computing, such as the quantum annealing procedure offered by D-Wave 
[S5; S40; S103]. This method utilises the quantum annealing algorithm for optimisation problems, with one of its most effective applications being quadratic unconstrained binary optimisation, which is useful for training support vector machines 
[S40]. Another approach is to employ gate-based quantum computers and leverage quantum algorithms, such as the Harrow-Hassidim-Lloyd algorithm, which can solve ordinary differential equations and perform matrix inversion in polylogarithmic time. This method accelerates the training process of classical neural networks that utilise the stochastic gradient descent algorithm to find optimal weights 
[S106]. The acceleration is achieved by pruning the neural network and encoding its weights into a sparse matrix, which is then loaded into a quantum circuit to perform matrix inversion in $\log(n)$ time, where 
$n$ is the number of weights.  Additionally, simulations indicate that quantum neural networks can be trained more quickly than their classical counterparts using stochastic gradient descent 
[S94]. Moreover, a quantum computer can achieve a quadratic speedup in training binary neural networks by utilising a quantum amplitude amplification 
[S110].

\noindent \textbf{Robustness against noise.} Another advantage offered by quantum machine learning is its robustness against noise 
[S4; S79; S93]. This robustness can be achieved in quantum generative adversarial models 
[S79]. Additionally, incorporating quantum layers into neural networks can lead to better performance than classical neural networks, as shown in simulation experiments for image classification problems under various types of noise 
[S4]. Furthermore, using multiple quantum layers sequentially, along with data reuploading, has outperformed several previously known quantum classifiers and some of the best classical counterparts, particularly in datasets with asymmetric Gaussian noise 
[S93]. 

\noindent \textbf{Robustness against adversarial attacks.} Robustness against well-engineered adversarial attacks is another potential advantage of quantum AI 
[S107]. With the development of quantum AI, researchers have raised concerns that inference engines could suffer from adversarial attacks similar to traditional neural networks 
[S25]. Indeed, it has been established in 
[S25] that quantum classifiers can be effectively deceived by minor adversarial perturbations. However, an adversarial training process, proposed by the authors, can significantly enhance their robustness against such attacks. The robustness of quantum neural networks against adversarial attacks has been confirmed in the context of fault classification for industrial control systems 
[S29] and image classification 
[S4, S8]. Additionally, it has been demonstrated that a quantum boost algorithm model trained with a quantum annealer is more robust to adversarial attacks compared to a state-of-the-art random forest model 
[S58]. General principles, perspectives, and challenges of quantum adversarial machine learning have been discussed in 
[S107]. In this work, the authors have established that the adversarial vulnerability of quantum classifiers originates from their underlying Hilbert spaces. They have noted several defence strategies for quantum classifiers, including adversarial robustness through quantum noise, certifiable robustness, and adversarial training. Finally in [S8], \cite{PhysRevResearch.5.023186} have benchmarked the robustness of several quantum variational classifiers and classical classifiers against  adversarial attacks. The results show that quantum classifiers offer a notably enhanced robustness against classical adversarial attacks by learning features, which are not detected by the classical neural networks, indicating a possible quantum advantage.

\noindent \textbf{Quantum sampling supremacy.} Quantum mechanical systems can produce probability distributions that exhibit quantum correlations, which are challenging to capture using classical models 
[S22]. Quantum generative models based on quantum neural networks facilitates efficient learning and sampling of generic probability distributions with an exponential speed-up compared to classical algorithms 
[S22; S71; S100]. While this capability may not have immediate practical applications at the model level,  it serves as a compelling demonstration of ``quantum learning supremacy,'' which holds significant interest for the advancement of quantum AI 
[S71; S100].

\section{Discussion}
\label{sec:trends}

\begin{figure}
    \centering
    \includegraphics[width=\linewidth]{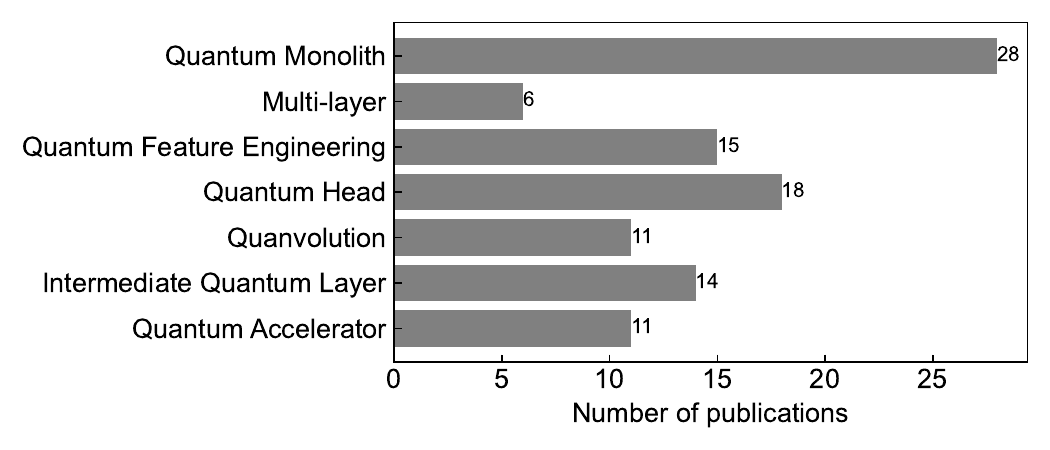}
    \caption{Number of publications related to each of the quantum-classic split patterns}
    \label{fig:bar-split}
\end{figure}

The number of publications corresponding to each of the architectural patterns described above is shown in Figure~\ref{fig:bar-split}. The most widely used pattern is the quantum monolith. This architecture is the simplest and is well-suited for demonstrating proof-of-concepts and conducting experiments in the field of QML. The next most used patterns are quantum head and intermediate quantum layer. These architectural solutions are closely tied to real-world AI applications for NISQ-era quantum computers. They are well-suited for high-dimensional data and offer promising advantages in the near future. 

Uploading large amounts of data directly into quantum computers poses challenges for NISQ-era systems due to limitations in the number of available qubits and their coherence time. As a result, the quantum feature extraction pattern is less commonly used. This pattern is mostly utilised in the quantum kernel methods, where the quantum component operates directly on data to compute the feature map. In \cite[]{Gil-Fuster_2024}, the universality -- and thus expressivity -- of embedding quantum kernels has been formally proven. The limitations related to a small number of qubits are somewhat alleviated by quantum convolutional filters, where the filter scans over the input data layer, processing only a small amount of data at a time. These systems are expected to significantly reduce the number of learnable parameters without loss of expressibility. However, this potential has not yet been fully realised on NISQ computers since it requires deep integration and communication between classical and quantum components.

The usage of the quantum accelerator pattern is also rare for practical applications in QML, since most quantum algorithms show substantial quantum advantage only in the limit of large data and require fault-tolerant computers, which is not achievable in NISQ devices.

The least usable pattern is the multi-layer approach. One of several implementations of this pattern is the quantum neural network architecture, where one layer is often considered a quantum neuron \cite[]{Tacchino2019}. Using several quantum layers might be considered a promising direction to improve the predictive accuracy of quantum ML, similar to classical deep learning, where the predictive power of the model can be enhanced with more layers of neurons \cite[]{DI9APDS7}. On the one hand, classical communication between layers reduces the depth of quantum circuits and lowers the requirements for qubit coherence time. On the other hand, the need for repetitive measurements and data re-uploading into the quantum computer may pose a bottleneck in terms of operational time and create an obstacle to achieving significant speed-up.

Note that existing alternatives to the quantum monolith architecture are primarily motivated by the limitations of the NISQ era, such as the limited number of quantum resources and noise. However, there are other fundamental challenges that may prevent the quantum monolith from being the optimal architecture, even when fault-tolerant, large-scale quantum computers become a reality. One such challenge is the trade-off between the expressivity and trainability of quantum neural networks, driven by the barren plateau problem. This issue arises due to the extremely high dimensionality of the underlying Hilbert space of quantum states. A more expressive quantum models also require a higher
measurement cost in gradient estimation \cite[]{chinzei2024}. While this dimensionality and expressivity can be reduced through specific quantum circuit designs, such as quantum convolutional neural networks (see Figure~\ref{fig:cirquit}d), circuits that avoid barren plateaus might be classically simulable, thereby negating the quantum advantage -- which is the primary motivation for the quantum monolith architecture \cite[]{cerezo2024, gilfuster2024}.

Below, we outline several trends that emerge in the development of novel quantum AI systems operating on NISQ devices, which we have identified in the literature.

\noindent \textbf{Scaling up.} Existing quantum software architectures are anticipated to scale up in the near future both in the number of qubits~\cite[]{QV4N5TVV, 5NIUVKSK, Y4B3NDVJ, RSDKE6EG, JA29UXLM} and the depth of quantum circuits~\cite[]{Tacchino2019, T4PG3ASU}. The reduced space and time complexity provided by quantum algorithms can offer practical advantages, but these benefits can be only observed with a large number of qubits. Consequently, scaling up is expected to improve the performance of current quantum software, allowing operations on higher-dimensional data. For instance, this will facilitate the processing of high-resolution graphic data with RGB channels, surpassing the limitations of grayscale processing~\cite[]{ULPQR2S6}. Another trend associated with scaling up is the increasing need to explore a wider variety of quantum circuit configurations and types~\cite[]{Huggins_2019, HRPF3VDM, 5NIUVKSK}.

\noindent \textbf{Architecture search and AutoQML.} Developing quantum software systems presents challenges due to the need for extensive interdisciplinary knowledge, the non-intuitive nature of quantum physics, and dependence on numerous parameters. Consequently, there is a growing trend to leverage automatisation and machine learning to develop quantum software. One of the practical implementations of this trend is automated quantum architecture search (QAS), which encompasses a set of computer-aided methods aimed at engineering optimal, task-specific parameterised quantum circuits~\cite[]{Zhang_2021}, which leads automated quantum machine learning (AutoQML) \cite[]{4HXFSHKM}. AutoQML consists of methods for automating the design and training of QML, reducing the need for human involvement \cite[]{4HXFSHKM, ALTARESLOPEZ2024122984, KoikeAkino2022AutoQMLAQ}.

Inspired by conventional neural architecture search, researchers employed an evolutionary search within a hierarchical representation of quantum circuits~\cite[]{Y523UU6T}, a method initially developed for deep neural networks in \cite[]{liu2018hierarchical}. Similarly, in~\cite[]{IS2X25QU}, other researchers created a search space from various configurations of entanglement layers represented by genotype vectors and used model-based optimization to identify the optimal configuration. In \cite[]{Zhang_2021}, supervised learning and neural networks were employed to learn quantum circuit architectures, while \cite[]{furrutter_quantum_2024} utilized generative machine learning models, specifically denoising diffusion models, to generate desired quantum operations within gate-based quantum circuits.

To enhance the efficiency of variational quantum algorithms, Du et al. introduced a one-stage optimization strategy~\cite[]{Du2022}. This approach optimizes both the parameterized network configuration, known as the supernet and its trainable parameters shared among sampled configurations. This method achieves the desired performance within reasonable runtime and memory usage constraints, demonstrating a practical solution for optimizing quantum circuits.

Further advancements in quantum architecture search include a cloud-based automated circuit architecture search framework proposed in~\cite[]{4HXFSHKM}. This framework enables parallelised hyperparameter exploration and model training, leveraging the power of cloud computing. Recognizing the critical role of benchmarking in advancing QAS, Lu et al. developed QAS-bench, a benchmark framework comprising two datasets with 900 quantum circuits and 400 unitary matrices, along with evaluation protocols to assess both existing and future QAS approaches~\cite[]{4NBYI7Z5}.

\noindent \textbf{Co-design.} The concept of software-hardware co-design is often used in the development of novel computing hardware that operates beyond binary logic~\cite[]{Klymenko}. The diversity of quantum hardware, the lack of a universal hardware architecture, and its continuous evolution contribute to its flexibility and allow for co-design approaches.  This co-design principle involves the joint optimization of software and hardware within specified constraints to ensure that applications are hardware-friendly and that hardware design is efficient~\cite[]{CHQ8NJHX}.

A practical application of this approach is mapping segments of classical neural architecture onto quantum circuits optimised for specific hardware, such as IBM Quantum Processors with superconducting qubits \cite[]{Jiang2021, BHB9VE94}. This mapping is effective when classical neural networks are designed appropriately. 

In~\cite[]{JKHCVY9X}, variational quantum circuits for quantum chemistry simulations are optimised considering hardware constraints. A class of these circuits, known as "hardware-efficient" ansatzes, can be executed on NISQ computers. The study also highlights the significance of application-specific quantum processor architectures for specialised tasks.

The hardware-efficient ansatz, designed for shallow quantum circuits, has been shown to support the construction of an equivariant quantum convolutional circuit architecture~\cite[]{PKIHPX3E}. Additionally, specific quantum hardware architectures exist for Pauli string manipulations in quantum chemistry simulations, enabling the execution of relevant quantum algorithms on current NISQ computers~\cite[]{CHQ8NJHX}.

\noindent \textbf{Distributed quantum computing} Distributed quantum computing  \cite[]{CALEFFI2024110672} presents an architecturally scalable solution for building large-scale quantum computing systems, addressing the limitations of NISQ systems. Several experiments have been conducted in this direction. Several authors \cite[]{5DKFHRBW, 3RITDWJQ} explored the distribution of partitioned features across multiple spatially separated quantum circuits (nodes) and reported a slight decrease in performance compared to a monolithic quantum neural network. While dataset distribution shares similarities with classical approaches, the distribution of quantum models is distinct due to the need for distributed quantum networks to establish entanglement across different nodes~\cite[]{IUEEV76R}. The inter-node two-qubit gates can be managed using three approaches: terminating two-qubit gates with measurements, approximating them with local gates, or establishing a coherent quantum communication channel between non-local qubits~\cite[]{IUEEV76R}.

Physically, the latter approach involves long-range communication with fibre optics to establish entangled Einstein-Podolsky-Rosen photon pairs~\cite[]{UTU2I6JY, 5DKFHRBW}. Communication between nodes cannot be broadband; thus, optimal partitioning of quantum circuits to minimise inter-node two-qubit gates is essential. This optimisation can be automated at the compiler level, as implemented in the QuComm framework~\cite[]{5DKFHRBW}, or manually using the Quantum Message Passing Interface, analogous to MPI functions in multi-processing programming~\cite[]{UTU2I6JY}. 

Additionally, the concept of federated learning, which has gained significant attention in classical deep learning, is considered beneficial for quantum machine learning~\cite[]{pmlr-v54-mcmahan17a, Li2021QuantumFL, 9842983, 9685012, 9746622}. It has been shown in \cite[]{heredge2024prospects} that QML with federated learning demonstrates the potential for increased privacy, unattainable in classical settings.

\noindent \textbf{DevOps.} A significant trend in quantum AI is the generation and deployment of quantum services through the adaptation of classic Development and Operations (DevOps) techniques~\cite[]{SN5TJDV6}. Implementing DevOps principles in the quantum domain involves automating the continuous integration and continuous delivery (CI/CD) pipelines for quantum software, ensuring that quantum algorithms can be efficiently updated, tested, and deployed. This approach accelerates the development lifecycle and enhances the reliability and stability of quantum software applications. Using continuous deployment and other DevOps techniques \cite[]{SN5TJDV6, D9TYIEZ6} can enhance the deployability of quantum software.

\section{Threats to validity}

Many papers on quantum software report results from experiments conducted on simulators rather than actual quantum computers. 
There is limited literature on QML in the context of hardware noise or errors, as well as on QML architectures incorporating quantum error correction. Additionally, most of the works considered in this review pertain to prototypes and experimental software systems that have not yet reached the production stage. The lack of real-world production experience in quantum software could influence existing architectural patterns in two ways: some patterns may not be adopted in production, while new patterns may emerge. There are only a few works that have conducted quantum validation on hardware, and the number of such studies is insufficient for a systematic mapping study. Nevertheless, the papers selected for this study offer valuable insights through theoretical reasoning, simulations, small-scale experiments, and toy models. These contributions allow us to identify potential architectural patterns for quantum AI systems even in the absence of full quantum validation. The theoretical justification for the usefulness of the identified patterns is strong enough to suggest their applicability in the production stage as well.

Another threat to validity is the lack of comprehensive systematic benchmarking of existing quantum software against state-of-the-art systems. Furthermore, establishing such benchmarking protocols is challenging due to the absence of standardized methodologies and metrics for comparing quantum ML systems. As a result, the published claims regarding the advantages of quantum systems may not hold up as strongly once proper benchmarking is conducted.

Another challenge discussed by \cite{Troyer2014} concerns the experimental evaluation of quantum speedup as reported in the literature. The authors argue that, to properly distinguish parallel speedup from quantum speedup, it is crucial to scale hardware resources identically for both quantum and classical devices during comparisons. Moreover, they emphasize the need to avoid confusing inefficiencies at small problem sizes with true indications of quantum speedup.

\section{Conclusions}

The unique software architectural solutions concerning quantum AI primarily revolve around determining which tasks to delegate to a quantum computer and how quantum and classical software components communicate with each other. During the systematic mapping study, we identified ten architectural patterns in quantum AI systems: seven related to the quantum-classic split and three related to quantum middleware. These patterns constitute the core of the pattern catalog developed in this work, which aims to guide software architects in effectively utilising quantum components in their projects. Each architectural pattern involves trade-offs in achieving the system quality attributes, such as efficiency, scalability, trainability, simplicity, portability, and deployability. The most commonly used patterns for small-scale proof-of-concept applications are the quantum monolith and quantum feature engineering. When attempting to scale the system up to handle larger volumes of information within the constraints of NISQ-era hardware, the corresponding challenges are typically addressed using architectural patterns such as the quantum head or quantum intermediate layer. In these architectures, part of the inference task, as well as the reduction of data dimensionality, is delegated to classical components. The primary approach for organising communication between classical and quantum components is through the use of a service-oriented architecture and service wrappers. In quantum AI systems, quantum components can be integrated into trainable inference models or function as accelerators for specific tasks.

The architectural patterns identified in this study describe architectural solutions that are specifically tailored for quantum AI systems. These patterns are crucial for addressing key issues such as computational efficiency, scalability and data flow management in those systems. By adopting these architectural patterns, developers can mitigate risks, improve maintainability, and optimize performance in quantum AI systems.

Quantum AI systems offer several benefits, such as accelerating the inference and training stages, enhancing robustness against noise in data and adversarial attacks, and achieving better accuracy. However, most of the collected evidence confirming these advantages is obtained from simulations on classical computers. There is no yet evidence that NISQ-era quantum computers can provide these benefits for practical applications. The identified trends in quantum AI are related to scaling up quantum components, automatic quantum architecture search, hardware-software co-design, utilising distributed quantum computing and federated learning, and developing quantum DevOps.

\bibliographystyle{elsarticle-harv} 
\bibliography{main}

\end{document}